\pdfoutput=1
\documentclass[12pt,english]{article}

\usepackage[T1]{fontenc}
\usepackage[utf8]{inputenc}
\usepackage{geometry}
\usepackage{amsmath}
\usepackage{amssymb}
\usepackage{graphicx}
\usepackage{setspace}
\makeatletter

\providecommand{\tabularnewline}{\\}

\newcommand{\lyxaddress}[1]{
	\par {\raggedright #1
	\vspace{1.4em}
	\noindent\par}
}

\usepackage[winfonts,UTF8]{ctex}

\usepackage{slashed}
\usepackage{hyperref}
\usepackage{graphicx}
\usepackage[numbers,sort&compress]{natbib}
\date{}

\hypersetup{colorlinks=true,citecolor=red,linkcolor=blue}
\usepackage{times}
\usepackage{amsfonts}

\makeatother

\usepackage{babel}
\begin{document}
\title{Type I critical dynamical scalarization and descalarization in Einstein-Maxwell-scalar
theory }
\author{Jia-Yan Jiang$^{1}$\thanks{jiangjy@stu2020.jnu.edu.cn}, Qian Chen$^{2}$\thanks{chenqian192@mails.ucas.ac.cn},
Yunqi Liu$^{3}$\thanks{yunqiliu@yzu.edu.cn}, Yu Tian$^{2,4}$\thanks{ytian@ucas.ac.cn},
Wei Xiong$^{5}$\thanks{phyxw@stu2019.jnu.edu.cn}, \\
Cheng-Yong Zhang$^{1}$\thanks{zhangcy@email.jnu.edu.cn, corresponding author}, Bin Wang$^{3,6}$\thanks{wang\_b@sjtu.edu.cn}}
\maketitle

\lyxaddress{\begin{center}
\textit{1. Department of Physics and Siyuan Laboratory, Jinan University,
Guangzhou 510632, China}\\
\textit{2. School of Physical Sciences, University of Chinese Academy
of Sciences, Beijing 100049, China}\\
\textit{3. Center for Gravitation and Cosmology, College of Physical
Science and Technology, Yangzhou University, Yangzhou 225009, China}\\
\textit{4. Institute of Theoretical Physics, Chinese Academy of Sciences,
Beijing 100190, China}\\
\textit{5. School of Physics and Optoelectronics, South China University
of Technology, Guangzhou 510641, China}\\
\textit{6. School of Aeronautics and Astronautics, Shanghai Jiao Tong
University, Shanghai 200240, China}
\par\end{center}}
\begin{abstract}
We investigated the critical dynamical scalarization and descalarization
of black holes within the framework of the Einstein-Maxwell-scalar
theory featuring higher-order coupling functions. Both the critical
scalarization and descalarization displayed first-order phase transitions.
When examining the nonlinear dynamics near the threshold, we always
observed critical solutions that are linearly unstable static scalarized
black holes. The critical dynamical scalarization and descalarization
share certain similarities with the type I critical gravitational
collapse. However, their initial configurations, critical solutions,
and final outcomes differ significantly. To provide further insights
into the dynamical results, we conducted a comparative analysis involving
static solutions and perturbative analysis.
\end{abstract}
\tableofcontents{}

\section{Introduction }

The discovery of critical gravitational collapse stands as one of
the most significant achievements in numerical relativity \cite{Choptuik:1992jv}.
This phenomenon arises when generic initial data for collapsing matter
is fine-tuned by adjusting a single parameter, denoted as $p$, to
reach the critical threshold $p_{*}$ for black hole formation. Under
these conditions, a unique linearly unstable critical solution (CS)
emerges. Acting as an attractor, this CS serves as a boundary, distinguishing
between two possible outcomes: the formation of a black hole spacetime
or the preservation of a flat spacetime. Notably, the CS can exhibit
characteristics of either a stationary star or a time-dependent self-similar
configuration, corresponding to type I or type II critical phenomena,
respectively \cite{Gundlach:2007gc}. In type II critical gravitational
collapse, which occurs within a localized region, the matter and metric
undergo pulsations on decreasing temporal and spatial scales until
a singularity forms at the center. Near the threshold, the resulting
black hole mass scales as $M\propto(p-p_{*})^{\gamma}$ in the vicinity
of the threshold. The critical exponent $\gamma$ is universal, independent
of the specific details of the initial data, although it does depend
on the nature of the collapsing matter \cite{Evans:1994pj,Abrahams:1993wa,Koike:1995jm,Gundlach:1995kd,Choptuik:2004ha}.
Type II critical gravitational collapse shares similarities with second-order
phase transitions, where the order parameter, in this case, the black
hole mass, exhibits continuity near the threshold. In contrast, type
I critical phenomena in gravitational collapse involve the formation
of a black hole with a minimum finite radius, resulting in a discontinuous
transition \cite{Brady:1997fj,Bizon:1998kq}. The duration of the
intermediate solution, which can be approximated by the CS at the
threshold, scales as $T\propto-\gamma\ln|p-p_{\ast}|$. The coefficient
$\gamma$ also exhibits universality in this context. Gravitational
collapse has been extensively studied in various scenarios, including
complex scalar fields \cite{Seidel:1991zh,Hawley:2000dt,Zhang:2015dwu},
Yang-Mills fields \cite{Bartnik:1988am,Bizon:1990sr,Gundlach:1996je,Choptuik:1996yg},
and modified gravity \cite{Liebling:1996dx,vanPutten:1996mt,Zhang:2016kzg}.
In asymptotically anti-de Sitter (AdS) spacetime, critical gravitational
collapse is connected to gravitational turbulent instability \cite{Bizon:2011gg,Dias:2011ss,Deppe:2014oua}. 

The gravitational collapse critical phenomena arise in the transition
from a flat space to a black hole. More recently, a new type of critical
gravitational dynamical behavior has been discovered in the context
of black hole scalarization within the framework of the Einstein-Maxwell-scalar
(EMS) theory featuring higher-order coupling functions \cite{Zhang:2021nnn}.
It appears during the transition from a metastable bald black hole
(BBH) to a metastable scalarized black hole (SBH). Here the BBH represents
the Reissner-Nordström (RN) black hole. Specifically, in the regime
of critical dynamical scalarization, when the perturbation parameter
$p$ approaches the threshold $p_{*}$, a CS which corresponds to
a linearly unstable SBH, always emerges. The intermediate solutions
persist on the CS for a duration scaling as $T\propto-\gamma\ln|p-p_{*}|$.
The final solutions near the threshold exhibit a mass gap. These behaviors
bear resemblance to type I critical gravitational collapse and are
referred to as type I critical dynamical scalarization. However, the
final masses of the resulting black holes follow power laws of the
form $M_{p}-M_{\pm}\propto|p-p_{*}|^{\gamma_{\pm}}$, where $M_{\pm}$
represent the masses of the SBH and BBH, respectively, as $p$ approaches
$p_{*}$ from above or below. The exponents $\gamma_{\pm}$ depend
on the initial data. These power laws are absent in type I critical
gravitational collapse and stem from the energy escaping to infinity.
In the case of asymptotically AdS spacetime, these power laws are
also absent due to energy confinement \cite{Zhang:2022cmu}. Furthermore,
in our work \cite{Zhang:2022cmu}, we explored the dynamical descalarization
and identified two distinct types of critical behaviors in the EMS
theory in asymptotically AdS spacetime. One type bears resemblance
to type I critical dynamical scalarization, where a linearly unstable
CS emerges. The other type is reminiscent of type II critical gravitational
collapse, characterized by a scalar value $\phi_{H}$ on the apparent
horizon following the relation $\phi_{H}\propto(p_{*}-p)^{1/2}$.
We refer to the latter as type II critical dynamical descalarization.
In a gravitational theory with scalar field coupling with both Gauss-Bonnet
invariant and Ricci scalar \cite{Liu:2022fxy}, we further found a
marginally stable CS in the type I critical dynamical descalarization. 

In this paper, we aim to expand upon our previous work presented in
\cite{Zhang:2021nnn}, which exclusively focused on the critical dynamical
scalarization occurring in asymptotically flat spacetime. Furthermore,
our previous study maintained a fixed total mass throughout the dynamical
evolution. However, this approach is somewhat unnatural as the total
mass should increase with the perturbation. In this current study,
we adopt a different approach by keeping the mass of the initial seed
RN black hole fixed and allowing the total mass to increase with the
perturbation. Through this modification, we discover the existence
of a threshold for descalarization, in addition to the threshold for
scalarization. Additionally, we explore the critical dynamical descalarization
starting from an initial seed SBH. All of these critical dynamics
fall under the category of type I. To enhance our understanding of
the dynamical outcomes, we conduct a comparative analysis with the
static solutions and perturbative analysis, thereby delving into the
intriguing distinctions between the type I critical dynamical scalarization
and type I gravitational collapse.

The organization of this paper is as follows. Section \ref{sec:EMS}
provides an introduction to the EMS theory. In Section \ref{sec:Dynamics},
we outline the numerical setups employed for the dynamical simulation
of scalarization and descalarization. We provide a comprehensive description
of the critical dynamics involved in these processes. Section \ref{sec:Static}
delves into the static solutions and quasinormal modes, enabling a
comparison with the dynamical results and offering valuable insights
into the critical dynamics. Finally, we conclude with a summary and
discussion in Section \ref{sec:Summary}.

\section{Einstein-Maxwell-scalar theory\label{sec:EMS}}

The action of the EMS theory considered in this study is given by
\begin{equation}
S=\frac{1}{16\pi}\int d^{4}x\sqrt{-g}\left[R-2\nabla_{\mu}\phi\nabla^{\mu}\phi-f(\phi)F_{\mu\nu}F^{\mu\nu}\right],
\end{equation}
where natural units are employed. Here, $R$ represents the Ricci
scalar, $F_{\mu\nu}=\partial_{\mu}A_{\nu}-\partial_{\nu}A_{\mu}$
denotes the field strength of Maxwell field $A_{\mu}$, and the real
scalar field $\phi$ nonminimally couples to the Maxwell invariant
through the coupling function $f(\phi)$. The Einstein equations are
given by
\begin{equation}
R_{\mu\nu}-\frac{1}{2}Rg_{\mu\nu}=2\left(T_{\mu\nu}^{\phi}+f(\phi)T_{\mu\nu}^{A}\right),
\end{equation}
where the energy-momentum tensor of the scalar and Maxwell fields
are expressed as
\begin{align}
T_{\mu\nu}^{\phi}= & \partial_{\mu}\phi\partial_{\nu}\phi-\frac{1}{2}g_{\mu\nu}\nabla_{\rho}\phi\nabla^{\rho}\phi,\\
T_{\mu\nu}^{A}= & F_{\mu\rho}F_{\nu}^{\ \rho}-\frac{1}{4}g_{\mu\nu}F_{\rho\sigma}F^{\rho\sigma}.\nonumber 
\end{align}
The equation of motion for the scalar field is given by
\begin{equation}
\nabla_{\mu}\nabla^{\mu}\phi=\frac{1}{4}\frac{df(\phi)}{d\phi}F_{\mu\nu}F^{\mu\nu},\label{eq:eqScalar}
\end{equation}
while the equations for the Maxwell field are
\begin{equation}
\nabla_{\mu}\left(f(\phi)F^{\mu\nu}\right)=0.\label{eq:eqMaxwell}
\end{equation}

In this study, we primarily focus on models characterized by a coupling
function of the form $f(\phi)=e^{\beta\phi^{n}}$, where $n$ is a
positive integer and $\beta$ denotes the coupling parameter that
quantifies the strength of the coupling. Specifically, when $n=1$,
these models are commonly known as Einstein-Maxwell-dilaton theories.
Such theories find relevance in low-energy string theories, supergravity
models, and Kaluza-Klein models \cite{Gibbons:1987ps,Garfinkle:1990qj}.
From (\ref{eq:eqScalar}), it becomes evident that in models with
$n=1$, RN black holes with $\phi=0$ do not constitute valid solutions,
leaving only the existence of dilatonic black holes. Dilatonic black
holes offer a theoretical framework for exploring the influence of
new degrees of freedom on the behavior of gravitational and electromagnetic
fields \cite{Ferrari:2000ep,Zhang:2015jda,Blazquez-Salcedo:2019nwd,Zhang:2021ybj}.
Furthermore, these black holes have been extensively studied in the
context of holographic models due to their intricate phase structures
\cite{DeWolfe:2011ts,Mo:2021jff,Zhang:2021edm,Zhao:2022uxc,Cai:2022omk}. 

In recent years, the EMS models with $n=2$ have attracted much attention
due to the phenomenon of spontaneous scalarization \cite{Herdeiro:2018wub,Fernandes:2019rez,Astefanesei:2019pfq,Guo:2021zed,Myung:2018jvi,Myung:2018vug,Myung:2019oua,Lai:2022ppn,Lai:2022spn,Wei:2022dzw,Lin:2023npr},
which has the potential to serve as a probe for testing the strong-field
regime of gravity and for explaining certain astrophysical observations
\cite{Martin:2007ue,Maleknejad:2012fw}. In this case, RN black holes
are solutions. But they may have tachyonic instability against scalar
perturbations when the coupling is strong and the charge to mass ratio
of the black hole is large. To be more concrete, the scalar perturbation
on the RN black hole background with a coupling function of $f(\phi)=e^{\beta\phi^{2}}$
is governed by the equation 
\begin{equation}
\nabla_{\mu}\nabla^{\mu}\delta\phi=\mu_{\text{eff}}^{2}\delta\phi,
\end{equation}
where the effective mass squared is given by $\mu_{\text{eff}}^{2}=-\frac{\beta Q^{2}}{2r^{4}}$.
Here $Q$ is the electric charge of the RN black hole. For positive
values of $\beta$, the scalar perturbation has a negative $\mu_{\text{eff}}^{2}$,
leading to a tachyonic instability. As a result, the linearly unstable
bald RN black holes will dynamically evolve into the linearly stable
SBHs \cite{Zhang:2021etr,Xiong:2022ozw,Luo:2022roz,Fernandes:2019rez,Herdeiro:2018wub}. 

More recently, there has been significant interest in studying models
with higher order coupling functions \cite{Blazquez-Salcedo:2020nhs,LuisBlazquez-Salcedo:2020rqp,Blazquez-Salcedo:2020crd}.
For instance, consider the case of $f(\phi)=e^{\beta\phi^{4}}$. In
this model, the RN black holes are stable and free of tachyonic instabilities,
but can be transformed into SBHs through a violent first-order dynamical
transition induced by large scalar perturbations. At the threshold
of perturbation, intriguing critical behaviors are observed \cite{Zhang:2021nnn,Zhang:2022cmu}.
This paper extends the work presented in \cite{Zhang:2021nnn} by
considering a scenario in which the total mass of the system increases
with the amplitude of the scalar perturbation. Interestingly, beyond
the threshold for scalarization, we also identify a threshold for
descalarization. We present static solutions and study their dominant
quasinormal modes, and compare them with those obtained from the dynamical
approach.

\section{Dynamical evolution\label{sec:Dynamics}}

In this section, we investigate the dynamical evolution of black holes
in EMS theories with the coupling function $f(\phi)=e^{\beta\phi^{4}}$.
We focus on two distinct types of initial configurations: (1) commencing
with a bald RN black hole and (2) initiating from a linealy stable
SBH. For both scenarios, we introduce an ingoing scalar perturbation
and simulate the evolution of the system. Nevertheless, it is important
to note that the qualitative behaviors observed during critical dynamical
scalarization and descalarization are not exclusive to these two specific
types of initial configurations.

\subsection{Equations for dynamical simulation}

To study the full nonlinear evolution of a black hole under large
perturbation in a spherically symmetric spacetime, we utilize the
Painlevé-Gullstrand (PG) coordinates, where the metric takes the form
\begin{equation}
ds^{2}=-\left(1-\zeta^{2}\right)\alpha^{2}dt^{2}+2\zeta\alpha dtdr+dr^{2}+r^{2}(d\theta^{2}+\sin^{2}\theta d\phi^{2}).\label{eq:PG}
\end{equation}
Here, $\zeta$ and $\alpha$ are metric functions that depend on $(t,r)$,
and the apparent horizon $r_{h}$ is located where $\zeta(t,r_{h})=1$.
The PG coordinates are regular on the apparent horizon and have been
used to study black hole dynamics \cite{Ziprick:2008cy,Kanai:2010ae,Ripley:2019tzx,Ripley:2019aqj,Ripley:2020vpk,Xiong:2022ozw,Zhang:2021nnn,Zhang:2021ybj,Liu:2022eri,Liu:2022fxy,Niu:2022zlf}.
For RN black hole solutions, $\alpha=1,\zeta=\sqrt{\frac{2M}{r}-\frac{Q^{2}}{r^{2}}}$
where $M$ is the total mass of the system and $Q$ the black hole
charge.

We take the Maxwell field as $A_{\mu}dx^{\mu}=A(t,r)dt.$ The Maxwell
equations (\ref{eq:eqMaxwell}) give 
\begin{equation}
\partial_{r}\left(r^{2}f(\phi)\frac{1}{\alpha}\partial_{r}A\right)=0,\ \ \ \partial_{t}\left(r^{2}f(\phi)\frac{1}{\alpha}\partial_{r}A\right)=0.
\end{equation}
These can be solved by 
\begin{equation}
\partial_{r}A=\frac{Q\alpha}{r^{2}f(\phi)}.
\end{equation}
Here $Q$ is an integration constant that interpreted as the black
hole electric charge. 

We introduce another two auxiliary variables 
\begin{align}
\Pi & =\frac{1}{\alpha}\partial_{t}\phi-\zeta\Phi.\label{eq:Pi}\\
\Phi & =\partial_{r}\phi,\label{eq:Phi}
\end{align}
The Einstein equations give 
\begin{align}
\partial_{r}\zeta= & \frac{r}{2\zeta}\left(\Phi^{2}+\Pi^{2}\right)+\frac{Q^{2}}{2r^{3}\zeta f(\phi)}+r\Pi\Phi-\frac{\zeta}{2r},\label{eq:zetadr}\\
\partial_{r}\alpha= & -\frac{r\Pi\Phi\alpha}{\zeta},\label{eq:alphadr}\\
\partial_{t}\zeta= & \frac{r\alpha}{\zeta}\left(\Pi+\Phi\zeta\right)\left(\Pi\zeta+\Phi\right).\label{eq:zetadt}
\end{align}
The equation of motion for the scalar field gives 
\begin{align}
\partial_{t}\phi & =\alpha\left(\Pi+\Phi\zeta\right),\label{eq:phit}\\
\partial_{t}\Pi & =\frac{\partial_{r}\left[\left(\Pi\zeta+\Phi\right)\alpha r^{2}\right]}{r^{2}}+\frac{\alpha}{2}\frac{Q^{2}}{r^{4}f^{2}(\phi)}\frac{df(\phi)}{d\phi},\label{eq:Pt}\\
\partial_{t}\Phi & =\partial_{r}\left[\alpha\left(\Pi+\Phi\zeta\right)\right].
\end{align}
We need to solve the metric functions $\alpha,\zeta$ and the scalar
field functions $\phi,\Phi,\Pi$. Given the initial scalar distribution
$\phi$ and $\Pi$, we can calculate the initial values of $\Phi,\zeta$
and $\alpha$ using equations (\ref{eq:Phi},\ref{eq:zetadr},\ref{eq:alphadr}),
respectively. Armed with these initial values, we can proceed to the
subsequent time slice, where we obtain the values of $\zeta,\phi,\Pi$
by employing the evolution equations (\ref{eq:zetadt},\ref{eq:phit},\ref{eq:Pt}),
respectively. The values of $\Phi$ and $\alpha$ can be obtained
by applying constraint equations (\ref{eq:Phi},\ref{eq:alphadr}),
respectively. By repeating this iterative procedure, we can obtain
all the metric and scalar functions at each time step. It is worth
noting that we only need to use the constraint equation (\ref{eq:zetadr})
once at the beginning.

\subsection{Numerical setup}

At large distances from the black hole, it can be demonstrated that
$\zeta=\sqrt{\frac{2M}{r}}\left(1+O(1/r)\right)$, where the constant
$M$ represents the total mass of the system, accounting for the energy
of the gravity, Maxwell and scalar fields \cite{Hayward:1994bu}.
Note that the Arnowitt-Deser-Misner (ADM) mass in PG coordinates always
evaluates to zero and does not reflect the correct physical mass of
the spacetime \cite{Shibata2016}. Therefore, we employ the Misner-Sharp
(MS) mass, defined as \cite{Misner:1964je} 
\begin{equation}
m(t,r)=\frac{r}{2}\left(1-g^{\mu\nu}\partial_{\mu}r\partial_{\nu}r\right)=\frac{r}{2}\zeta(t,r)^{2}.\label{eq:ms}
\end{equation}
It can be easily shown that the RN black hole possesses $m=M-\frac{Q^{2}}{2r}$.
The MS mass can be regarded as the radially integrated energy density
of the energy momentum tensor \cite{Ripley:2019irj,Ripley:2019aqj}.
On the apparent horizon $r_{h}$, we have $m(t,r_{h})=\frac{1}{2}r_{h}$,
which equals to the irreducible mass $M_{h}$ of the black hole defined
by $M_{h}=\sqrt{\frac{A_{h}}{16\pi}}$, where $A_{h}=4\pi r_{h}^{2}$
denotes the area of the apparent horizon. The total mass of the system
can then be calculated as 
\begin{equation}
M=\lim_{r\to\infty}m(t,r)=\lim_{r\to\infty}\frac{r}{2}\zeta(t,r)^{2}.\label{eq:M}
\end{equation}

To ensure the stability and long-term evolution of the numerical simulation,
we encounter challenges stemming from the decay of $\zeta$. As a
remedy, we introduce a new variable $s=\sqrt{r}\zeta$ in our simulations.
Additionally, the decay of $\zeta$ poses difficulties in imposing
outer boundary conditions at a fixed finite $r$ \cite{Ripley:2019aqj}.
To address this issue, we introduce a coordinate compactification
given by 
\begin{equation}
z=\frac{r}{1+r/L},
\end{equation}
where $L=1$ represents a fixed unit length scale. The system is evolved
within the region $z\in[z_{c},1)$, where $z=1$ corresponds to the
spatial infinity. Here $z_{c}=\frac{r_{c}}{1+r_{c}}$, with $r_{c}$
denoting a cutoff located close to the initial apparent horizon $r_{h}$
from the interior. Consequently, we evolve the system within the region
$r\in[r_{c},\infty)$. It should be noted that since the radius of
apparent horizon never decreases in EMS theory, $r_{c}$ always remains
inside the apparent horizon throughout the entire evolution. We discretize
$z$ uniformly using $2^{12}$ grid points. The resolution is limited
in the far region during late times, but this does not affect our
analysis as our focus is on the near horizon behavior. We employ the
finite difference method in the radial direction and the fourth-order
Runge-Kutta method in the time direction. To stabilize the numerical
simulation, we utilize the Kreiss-Oliger dissipation. In the first
step, we apply the Newton-Raphson method to solve the constraint equation
(\ref{eq:zetadr}), where the L'Hospital's rule is used at $z=1$. 

We have checked the accuracy and convergence of our numerical method
by various ways. The convergence of finite difference method is often
estimated by $\frac{u_{2N}-u_{N}}{u_{4N}-u_{2N}}=2^{n}+O(\frac{1}{N})$
in which $u_{N}$ is the results by using $N$ grid points, $n$ is
the accurate order. It turns out that our numerical solutions converge
indeed to fourth order \cite{Zhang:2021ybj,Zhang:2021nnn,Liu:2022fxy}.
Furthermore, we have also employed the second-order finite difference
method to simulate the evolution and the results are qualitatively
consistent.

\subsubsection{Initial conditions}

We take two distinct types of initial configuration:

(1) The seed black hole is a BBH with $M_{0}=1,Q=0.9.$ It has metric
function $\zeta_{0}(r)=\sqrt{\frac{2M_{0}}{r}-\frac{Q^{2}}{r^{2}}}$
and scalar distribution $\phi_{0}(r)=0$. The initial apparent horizon
locates at $r_{h}=1.436$ or $z_{h}=0.5895$. We take an ingoing initial
scalar perturbation with the form 
\begin{align}
\delta\phi(r)= & p\begin{cases}
10^{-2}e^{-\frac{1}{r-r_{1}}-\frac{1}{r_{2}-r}}(r_{2}-r)^{2}(r-r_{1})^{2}, & r_{1}<r<r_{2},\\
0, & \text{otherwise},
\end{cases}\label{eq:initialRN}\\
\Pi(r)= & \partial_{r}\delta\phi(r).\nonumber 
\end{align}
in which $r_{1}=4,r_{2}=9$ and $p$ is the perturbation amplitude
parameter. 

(2) The seed black hole is an SBH with $M_{0}=1.2,Q=0.9$ and the
scalar charge $Q_{s}=0.6939$$.$ It is obtained by solving the static
equations of motion directly, as will be described in detail in subsection
\ref{subsec:Static-solutions}. We still refer to the metric function
of the seed black hole as $\zeta_{0}(r)$ and nontrivial scalar distribution
as $\phi_{0}(r)$. The initial apparent horizon is located at $r_{h}=2.149$
or $z_{h}=0.6824$. We introduce an ingoing scalar perturbation with
the form 
\begin{align}
\delta\phi(r)= & p\begin{cases}
10^{-2}e^{-\frac{1}{r-r_{1}}-\frac{1}{r_{2}-r}}(r_{2}-r)^{2}(r-r_{1})^{2}, & r_{1}<r<r_{2},\\
0, & \text{otherwise},
\end{cases}\label{eq:initialSBH}\\
\Pi(r)= & \partial_{r}\delta\phi(r)-\zeta_{0}(r)\partial_{r}\phi_{0}(r).\nonumber 
\end{align}
in which $r_{1}=6,r_{2}=10$ and $p$ is the perturbation amplitude
parameter. 

\subsubsection{Boundary conditions }

We specify following boundary condition for (\ref{eq:alphadr}) to
solve $\alpha$ in the evolution:
\begin{equation}
\alpha|_{r\to\infty}=1,\label{eq:alphab}
\end{equation}
This choice of boundary condition stems from the auxiliary freedom
of $\alpha dt$ in PG coordinates and implies that an observer at
infinity will measure time using the proper time coordinate $t$.
We also requre that 
\begin{equation}
\phi|_{r\to\infty}=0,\ \ \ \Pi|_{r\to\infty}=0,\ \ \ \Phi|_{r\to\infty}=0,
\end{equation}
during the evolution. They imply that $\partial_{t}\phi|_{r\to\infty}=\partial_{t}\Pi|_{r\to\infty}=0$
in (\ref{eq:phit},\ref{eq:Pt}). These conditions are sensible as
matter cannot reach spatial infinity in a finite amount of time. 

As previously mentioned, we use (\ref{eq:zetadr}) to solve for the
initial values of $\zeta$ or $s$. For this purpose, we specify the
following boundary condition: 
\begin{equation}
\zeta|_{t=0,r=r_{c}}=\zeta_{0}(r_{c}).
\end{equation}
Here $\zeta_{0}(r)$ refers to the metric function of the initial
seed BBH or SBH, as described in the preceding subsection. This boundary
condition is reasonable since the initial perturbation affects only
the geometry in $r>r_{2}$ in both cases, while the geometry in $r\le r_{1}$
remains unchanged by the perturbation at the initial time. This aspect
will be explicitly demonstrated in Fig.\ref{fig:RN0MS} and Fig.\ref{fig:SBH0MS}.
After obtaining the initial $\zeta$ or $s$ using (\ref{eq:zetadr}),
we can calculate the total mass of the system via (\ref{eq:M}). During
subsequent evolution, we fix 
\begin{equation}
s|_{r\to\infty}=\sqrt{2M}.
\end{equation}
This specification implies that we set $\partial_{t}s|_{r\to\infty}=0$
in (\ref{eq:zetadt}) during the evolution. 

\subsection{Dynamical results when the seed black hole is a RN black hole \label{subsec:DynamicalRN}}

\begin{figure}[h]
\begin{centering}
\begin{tabular}{cc}
\includegraphics[width=0.48\linewidth]{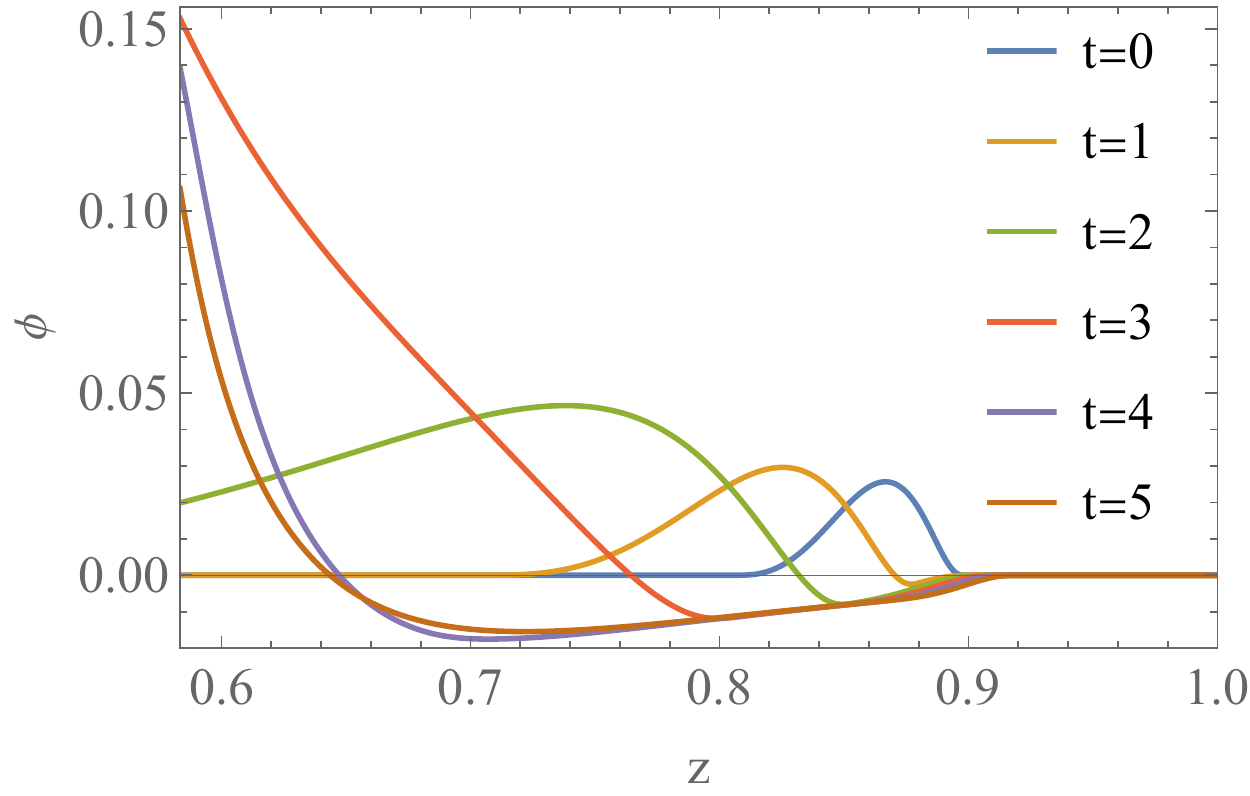} & \includegraphics[width=0.48\linewidth]{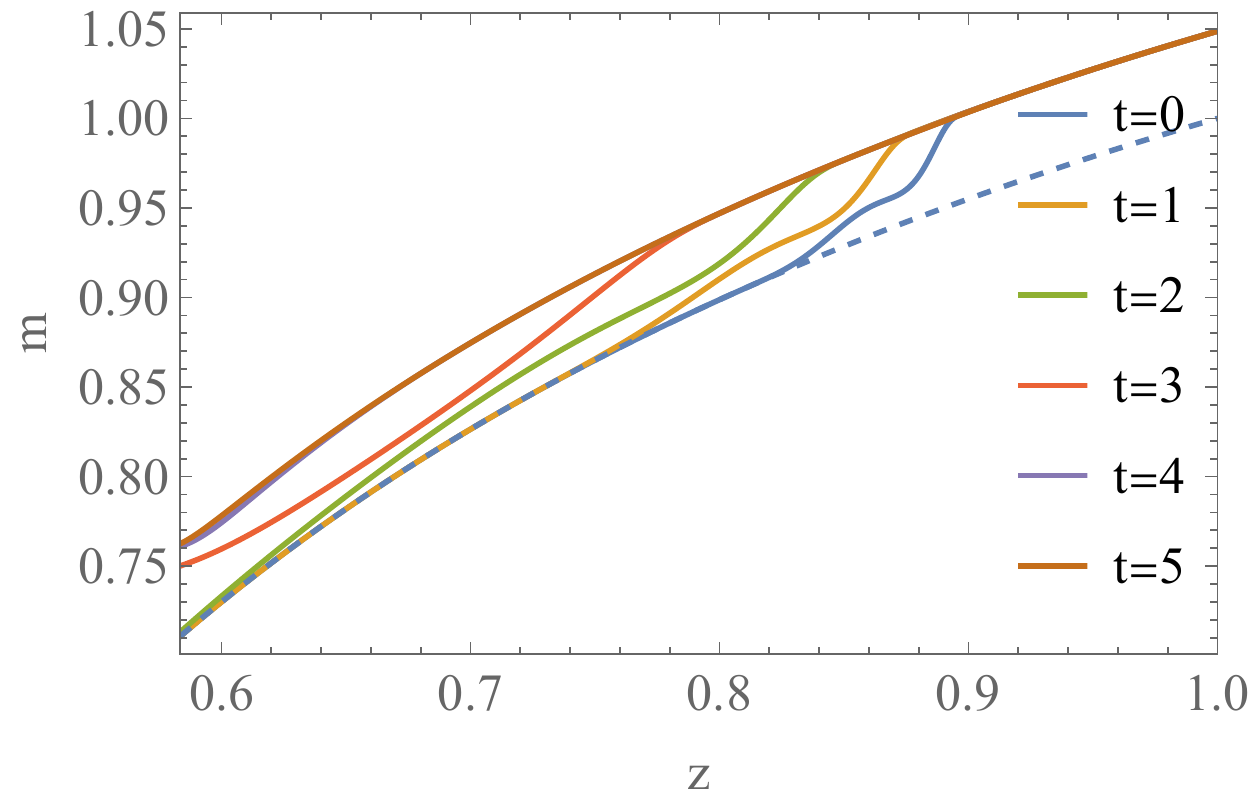}\tabularnewline
\end{tabular}
\par\end{centering}
{\footnotesize{}\caption{{\footnotesize{}\label{fig:RN0MS}The early evolution of the scalar
field and MS mass when the seed black hole is a bald RN black hole
with a total mass $M=1$. In the right panel, the dashed blue curve
represents the MS mass for the seed RN black hole. The solid blue
curve with $t=0$ represents the MS mass of a nonequilibrium spacetime
when the ingoing perturbation (\ref{eq:initialRN}) with amplitude
$p=0.1462$ is applied. The total mass increases to $M=1.049$. }}
}{\footnotesize\par}
\end{figure}

We fix the coupling parameter $\beta=2000$ and the black hole chagre
$Q=0.9$ in this subsection. The seed black hole is a BBH with a total
mass $M=1$. After applying the ingoing scalar field perturbation
(\ref{eq:initialRN}), the spacetime transforms into a nonequilibrium
state and the dynamical evolution commences. In Fig.\ref{fig:RN0MS},
we present the early evolution of the scalar field and MS mass. The
scalar perturbation propagates inwards, and drives the evolution of
the seed black hole. As implied by the right panel, in the beginning,
the spacetime geometry remains unchanged in the region $r<r_{1}$,
resembling that of the seed RN black hole. However, in the region
$r>r_{2}$, the geometry resembles that of a larger RN black hole
with a larger total mass, where the increase in mass arises from the
perturbation of the scalar field. 

Within the family of initial data (\ref{eq:initialRN}) parametrized
by $p$, we observe the presence of two distinct thresholds: $p_{1}\approx0.1461857045705$
and $p_{2}\approx0.2789936530457$. It is worth noting that while
the last few digits of these threshold values may be influenced by
numerical specifics, the first few digits remain consistent. When
$p$ is below $p_{1}$, the final black hole remains a BBH. Conversely,
when $p$ lies between $p_{1}$ and $p_{2}$, the final black hole
undergoes scalarization and becomes an SBH. Interestingly, for $p$
exceeding $p_{2}$, the final black hole reverts to being a BBH once
again. As such, we refer to $p_{1}$ and $p_{2}$ as the thresholds
for dynamical scalarization and descalarization, respectively.

\subsubsection{Dynamical critical behaviors of scalarization}

\begin{figure}[h]
\begin{centering}
\begin{tabular}{cc}
\includegraphics[width=0.48\linewidth]{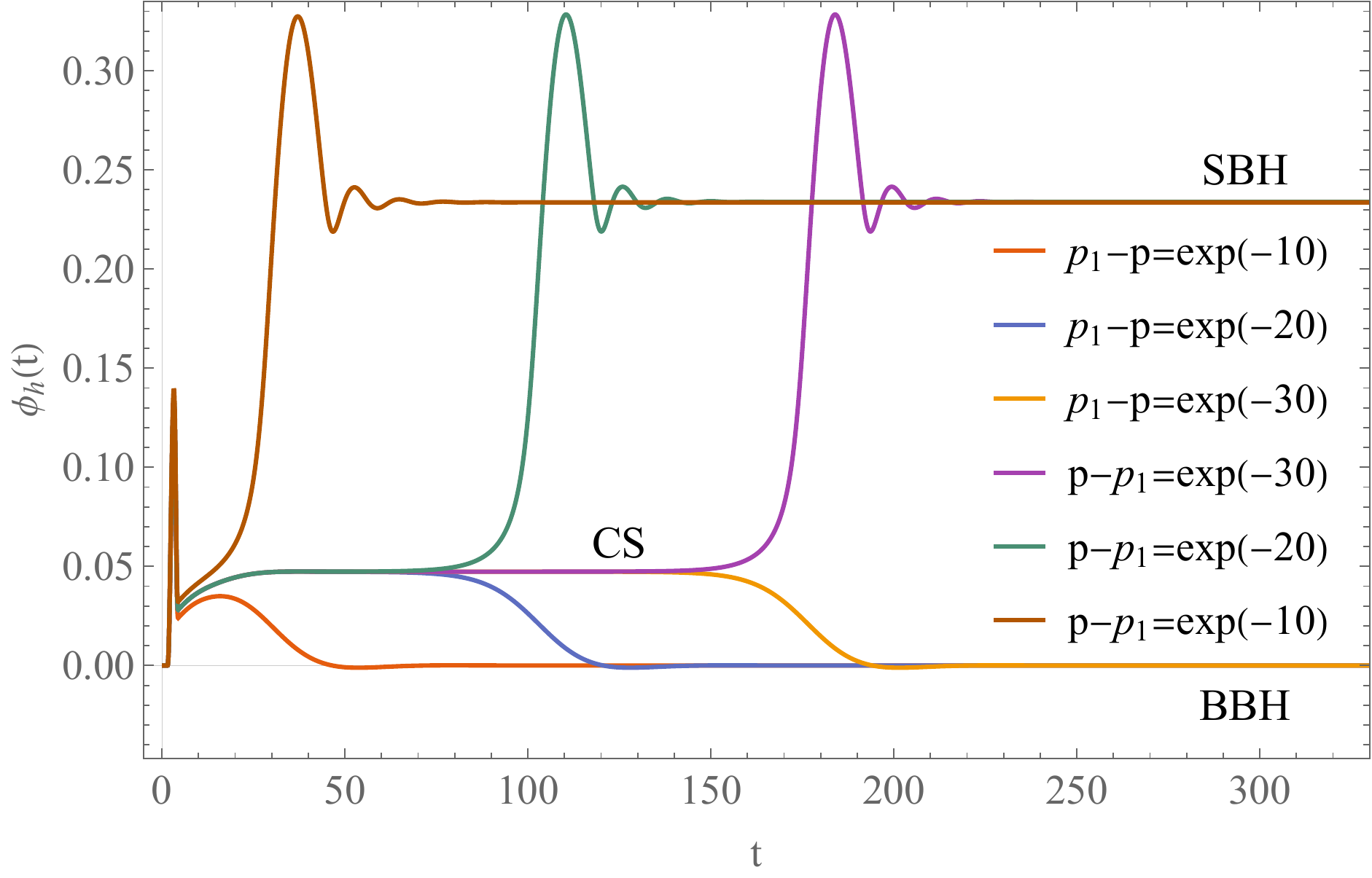} & \includegraphics[width=0.48\linewidth]{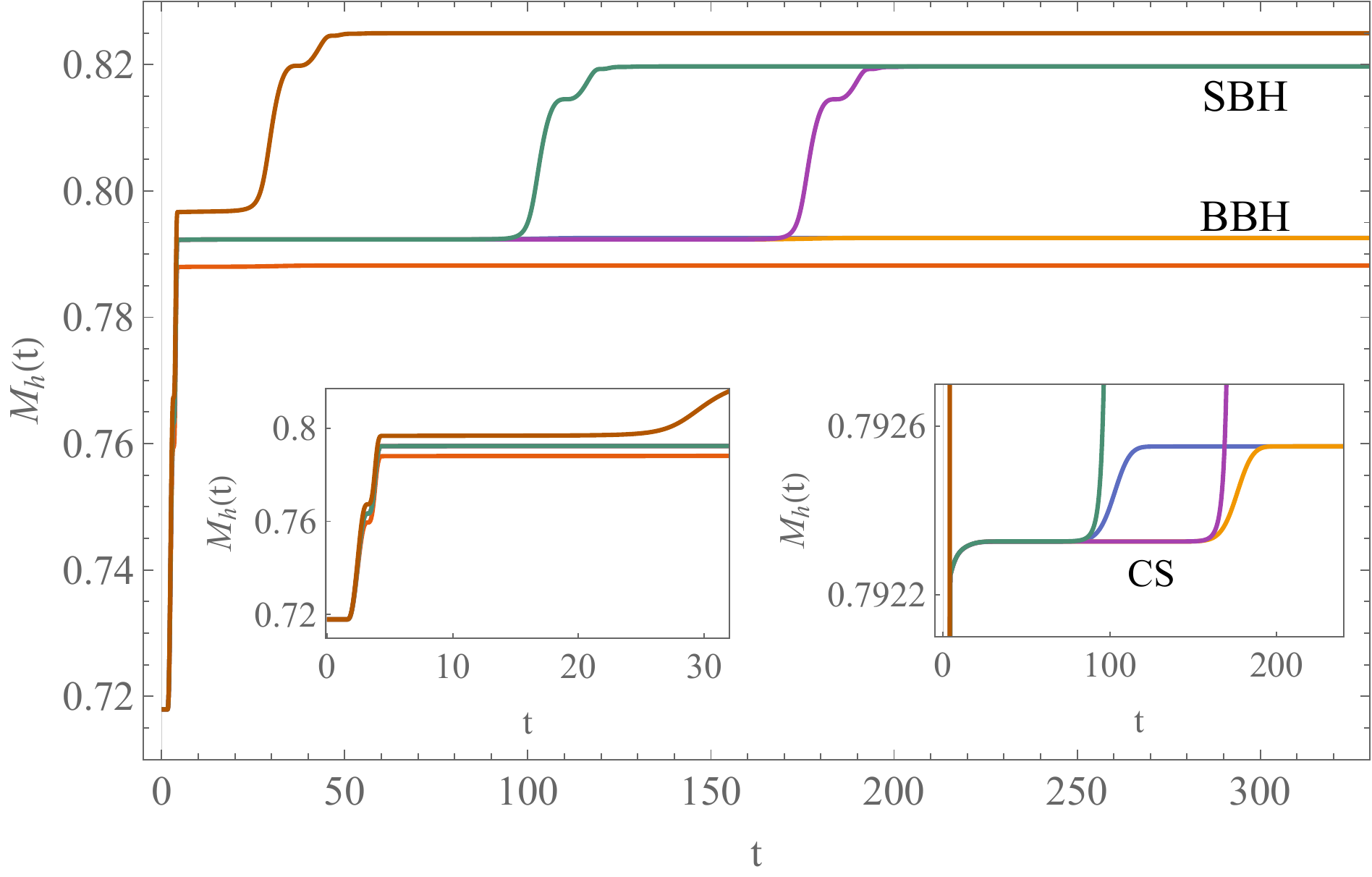}\tabularnewline
\end{tabular}
\par\end{centering}
{\footnotesize{}\caption{{\footnotesize{}\label{fig:RN0Mphihtp1}The evolution of the scalar
field on the apparent horizon $\phi_{h}$ and the black hole irreducible
mass $M_{h}$ with various $p$ near $p_{1}\approx0.1461857045705$
for initial data family (\ref{eq:initialRN}). The irreducible mass
of the CS is very close to that of the final BBH. Their differences
are shown in the right inset in the right panel. The left inset shows
the evolution of $M_{h}$ in the early times. }}
}{\footnotesize\par}
\end{figure}

Let us consider the dynamics for scalarization near $p_{1}$ at first.
We monitor the evolution of the scalar value $\phi_{h}$ on the apparent
horizon and the black hole irreducible mass $M_{h}$. The results
are depicted in Fig.\ref{fig:RN0Mphihtp1}. From the left panel, we
see that $\phi_{h}$ evolves from zero. As $p$ approaches $p_{1}$
from either below or above, after experiencing a rapid change in the
early stages of evolution, all the intermediate solutions are attracted
to a plateau. The closer $p$ is to $p_{1}$, the longer $\phi_{h}$
remains on this plateau. Essentially, the plateau represents a linearly
unstable static CS. By precisely fine-tuning $p$ to the exact threshold
$p_{1}$, the evolution would hypothetically remain indefinitely on
the CS. However, at late times, the evolution of $\phi_{h}$ deviates
from the CS. As long as $p<p_{1}$, the final black hole remains bald,
while if $p>p_{1}$, the black hole undergoes scalarization, acquiring
a nonvanishing scalar field and transforming into an SBH. We use the
term subcritical to describe the case where $p<p_{1}$, critical to
describe the case where $p=p_{1}$, and supercritical to describe
the case where $p>p_{1}$.

The phenomenon of a plateau is also observed in the right panel of
Fig.\ref{fig:RN0Mphihtp1}. The duration of $M_{h}$ staying on this
plateau becomes longer as $p$ approaches $p_{1}$. It is worth noting
that the irreducible mass of the black hole never decreases during
the evolution, which aligns with the expectation based on the second
law of black hole thermodynamics. In the subcritical case, the final
BBH has a irreducible mass very close to that of the CS. In the supercritical
case, the final SBH has a much larger irreducible mass than the CS. 

In summary, the evolution process for $p$ near threshold $p_{1}$
can be summerized as follows:
\begin{equation}
\text{BBH (metastable) }+\text{perturbation}\to\begin{cases}
\text{BBH (metastable)}, & \text{subcritical }(p<p_{1}),\\
\text{CS (unstable SBH)}, & \text{critical }(p=p_{1}),\\
\text{SBH (metastable)}, & \text{supercritical }(p_{1}<p<p_{2}).
\end{cases}\label{eq:p1CS}
\end{equation}
The initial seed RN black hole behaves as a lineary stable BBH under
small scalar field perturbation. However, when the perturbation becomes
sufficiently strong, scalarization occurs and the BBH transitions
into a linearly stable SBH. At the precise threshold, a CS arises,
which represents a linearly unstable SBH and acts as an attractor. 

\begin{figure}[h]
\begin{centering}
\includegraphics[width=0.5\linewidth]{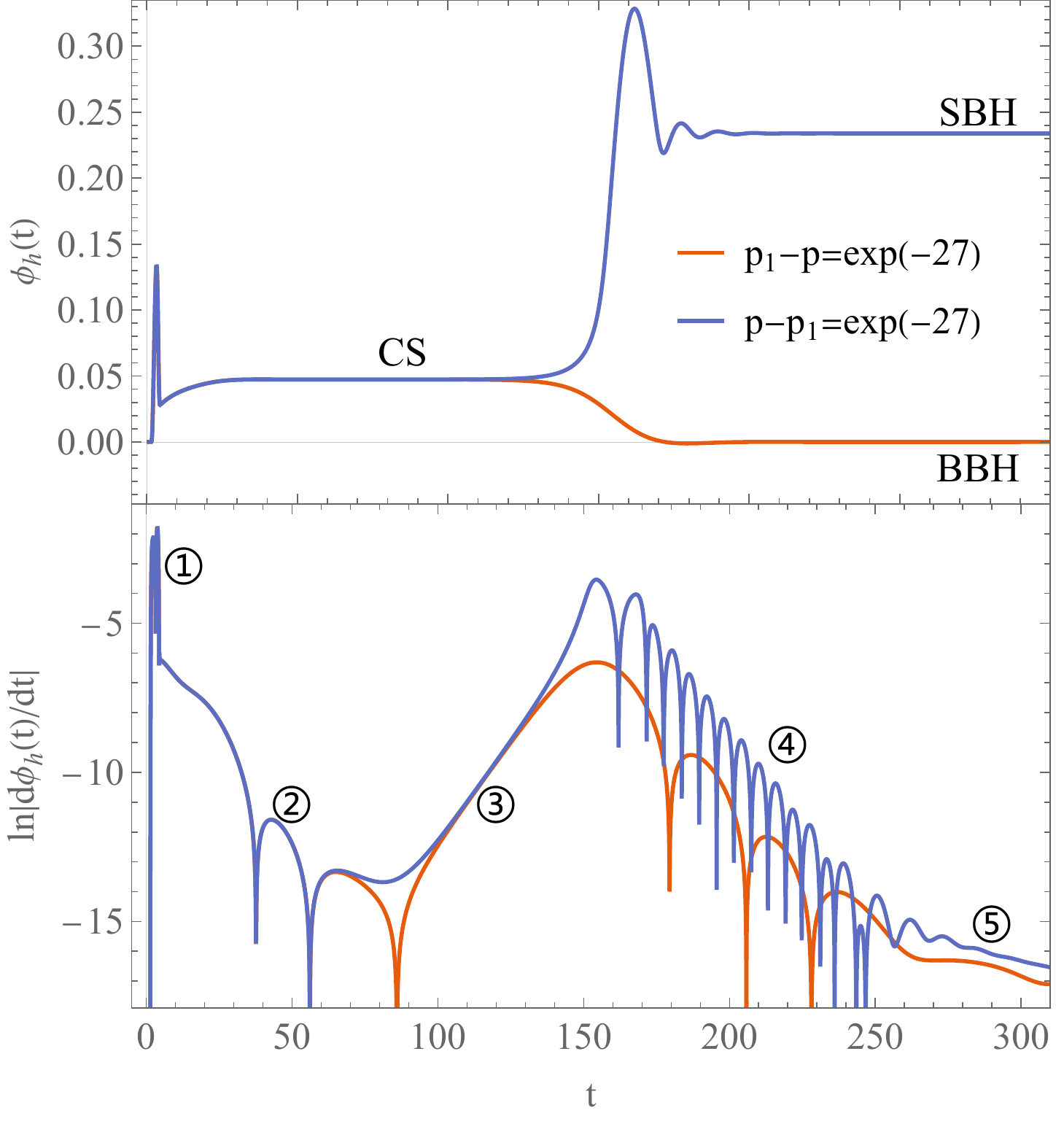}
\par\end{centering}
{\footnotesize{}\caption{{\footnotesize{}\label{fig:RN0phihdtlog}The evolution of $\phi_{h}$
and $\ln|\frac{d\phi_{h}}{dt}|$ when $p-p_{1}=\pm e^{-27}$ for initial
data (\ref{eq:initialRN}).}}
}{\footnotesize\par}
\end{figure}

To provide a more detailed analysis of the dynamical process, we present
the evolution of $\ln|\frac{d\phi_{h}}{dt}|$ in Fig.\ref{fig:RN0phihdtlog}.
Regardless of whether the case is subcritical or supercritical, the
entire evolution can be divided into five distinct stages. In the
first stage, the incoming scalar field perturbation is captured by
the seed black hole, resulting in a sudden increase in the irreducible
mass of the black hole, as depicted in the right panel of Fig.\ref{fig:RN0Mphihtp1}.
The scalar value $\phi_{h}$ on the apparent horizon also undergoes
drastic changes. The second and third stages, as shown in Fig.\ref{fig:RN0phihdtlog},
correspond to the plateau. In both subcritical and supercritical cases,
the intermediate solution converges to the CS with a damping rate
$\nu_{1}\approx0.14$ in the second stage, while subsequently departing
exponentially from the CS with an exponent $\eta_{1}\approx0.136$
in the third stage. The fourth and fifth stage can be interpreted
as the quasinormal modes (QNMs) and late time tail of the final black
hole, respectively, under small perturbation. The dominant mode can
be extracted by using the Prony method \cite{Berti:2007dg}. For the
supercritical case, the dominant mode exhibits a complex frequency
$\omega_{1s}\approx0.532-0.126i$, while for the subcritical case,
the dominant mode has a frequency $\omega_{1b}\approx0.115-0.099i$.

\begin{figure}[h]
\begin{centering}
\includegraphics[width=0.5\linewidth]{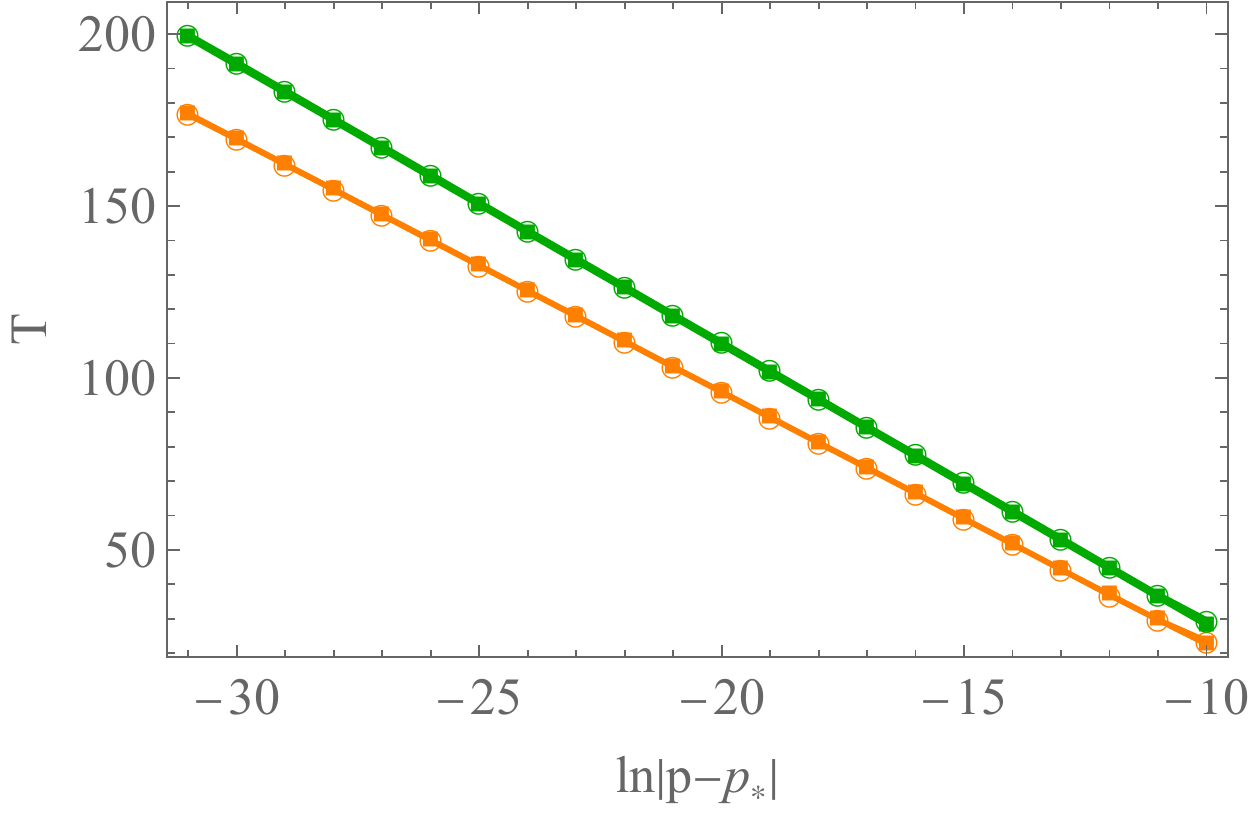}
\par\end{centering}
{\footnotesize{}\caption{{\footnotesize{}\label{fig:RN0Tlnp}The time $T$ of the intermediate
solution remains on the CS with respect to $\ln|p-p_{*}|$ for initial
data (\ref{eq:initialRN}). Here $p_{*}$ represents thresholds $p_{1}$
or $p_{2}$. The orange circles and squares are results for subcritical
and supercritical cases near $p_{1}$, respectively. The green circles
and squares are results for subcritical and supercritical cases near
$p_{2}$, respectively.}}
}{\footnotesize\par}
\end{figure}

The duration $T$ for which the intermediate solution remains on the
CS can be estimated by observing the time when $\ln|\frac{d\phi_{h}}{dt}|$
reaches its maximum after the first stage. This maximum occurs at
the turning point between the third and fourth stages. For values
of $p$ that are close to $p_{1}$, we have found 
\begin{equation}
T\propto-\gamma_{1}\ln|p-p_{1}|,\label{eq:Tlnp1}
\end{equation}
where $\gamma_{1}=\frac{1}{\eta_{1}}\approx7.36$. This relationship
is explicitly shown in Fig.\ref{fig:RN0Tlnp}. It suggests that the
intermediate solution in the second and third stages can be approximated
as 
\begin{equation}
\phi_{p_{1}}(t,r)=\phi_{*1}(r)+(p-p_{1})e^{\eta_{1}t}\delta\phi_{1}(r)+\text{stable modes.}\label{eq:approx}
\end{equation}
Here $\phi_{*1}(r)$ represents the CS at the exact threshold $p_{1}$,
and $\eta_{1}$ is the eigenvalue associated with the single unstable
eigenmode $\delta\phi_{1}(r)$. The stable modes dominates the evolution
during the second stage. In the third stage, the unstable mode $\delta\phi_{1}(r)$
starts to dominate as its coefficient grows to a finite size, namely,
$|p-p_{1}|e^{\eta_{1}T}\sim O(1)$ which implies (\ref{eq:Tlnp1}).
When $t\approx T$, the unstable mode grows to the same order as $\phi_{*1}(r)$.
The backreaction of the unstable modes on the spacetime destroys the
CS, and leads to the transition of the CS into a linearly stable BBH
with a vanishing scalar field when $p<p_{1}$, or a linearly stable
SBH with a nonvanishing scalar field when $p>p_{1}$. It is worth
noting that the approximation (\ref{eq:approx}) bears resemblance
to the one found in type I critical gravitational collapse \cite{Evans:1994pj,Koike:1995jm,Bizon:1998kq}.
It has been also found in sGB theory \cite{Liu:2022fxy} and some
holographic models \cite{Chen:2022cwi,Chen:2022vag,Chen:2022tfy}. 

We have also investigated the evolution of $\ln\frac{dM_{h}}{dt}$,
and discovered an intriguing relationship that holds throughout most
of the dynamic process:
\begin{equation}
\ln\frac{dM_{h}}{dt}=2\ln\left|\frac{d\phi_{h}}{dt}\right|+\text{const}.\label{eq:twice}
\end{equation}
This relation remains valid for various initial amplitudes $p$ and
system parameters like $M_{0},Q,\beta$. Note that the black hole
irreducible mass never decreases, so there is no need to take the
absolute value of $\frac{dM_{h}}{dt}$ in (\ref{eq:twice}). This
relation can be understood through perturbative analysis of static
background solutions. Suppose $\alpha_{0}(r),\zeta_{0}(r)$ are background
metric functions, and $\Pi_{0}(r),\Phi_{0}(r),\phi_{0}(r)$ are background
scalar field functions. As we have mentioned before (\ref{eq:M}),
the irreducible mass is equal to half of the black hole horizon areal
radius $r_{h}$. Using the fact that the apparent horizon is located
where $\zeta(t,r_{h})=1$, its evolution can be expressed as 
\begin{equation}
\frac{dM_{h}}{dt}=\frac{1}{2}\frac{dr_{h}}{dt}=-\frac{1}{2}\left.\frac{\partial_{t}\zeta}{\partial_{r}\zeta}\right|_{r=r_{h}},
\end{equation}
For static solutions, we have $\left.\zeta_{0}\right|_{r=r_{h}}=1$.
On the horizon, (\ref{eq:zetadt}) gives $\left.\Pi_{0}+\Phi_{0}\right|_{r=r_{h}}=0$.
Then the leading order of (\ref{eq:Pi}) gives
\begin{equation}
\left.\delta\Pi\right|_{r=r_{h}}=\left.\frac{1}{\alpha_{0}}\partial_{t}\delta\phi-\delta\Phi\right|_{r=r_{h}}.
\end{equation}
Here $\delta\Pi,\delta\Phi,\delta\phi$ represent perturbations on
the static background. The leading order of (\ref{eq:zetadt}) gives
\begin{equation}
\left.\partial_{t}\delta\zeta\right|_{r=r_{h}}=\left.r\alpha_{0}\left(\delta\Pi+\delta\Phi\right)^{2}\right|_{r=r_{h}}=\frac{r_{h}}{\alpha_{h}}\left.(\partial_{t}\delta\phi)^{2}\right|_{r=r_{h}},
\end{equation}
where $\alpha_{h}=\alpha_{0}(r_{h})$. By combining the expansion
(\ref{eq:bdz}) and temperature (\ref{eq:Temperature}) for static
solutions, we obtain 
\begin{equation}
\frac{dM_{h}}{dt}=\frac{r_{h}^{2}}{\alpha_{h}}\left(1-\frac{Q^{2}}{f(\phi_{h})r_{h}^{2}}\right)^{-1}\left.(\partial_{t}\delta\phi)^{2}\right|_{r=r_{h}}=\frac{r_{h}}{4\pi T_{h}}\left.(\partial_{t}\delta\phi)^{2}\right|_{r=r_{h}}.
\end{equation}
Taking the logarithm of the equation, we arrive at equation (\ref{eq:twice}).
The const term in (\ref{eq:twice}) is equal to $\ln\frac{r_{h}}{4\pi T_{h}}$.
It differs between the CS and the final black hole with or without
a scalar field (BBH or SBH). This relationship emphasizes that the
second and third stages can be viewed as perturbations on the CS,
while the fourth and fifth stages can be seen as perturbations on
the final BBH or SBH. It represents an enhanced version of the similar
relationships found in other EMS models \cite{Zhang:2021etr,Zhang:2021edm,Zhang:2021ybj,Zhang:2021nnn}.
But this relation may not hold in other theory, for example, the sGB
theory \cite{Liu:2022eri,Liu:2022fxy}.

\begin{figure}[h]
\begin{centering}
\begin{tabular}{cc}
\includegraphics[width=0.48\linewidth]{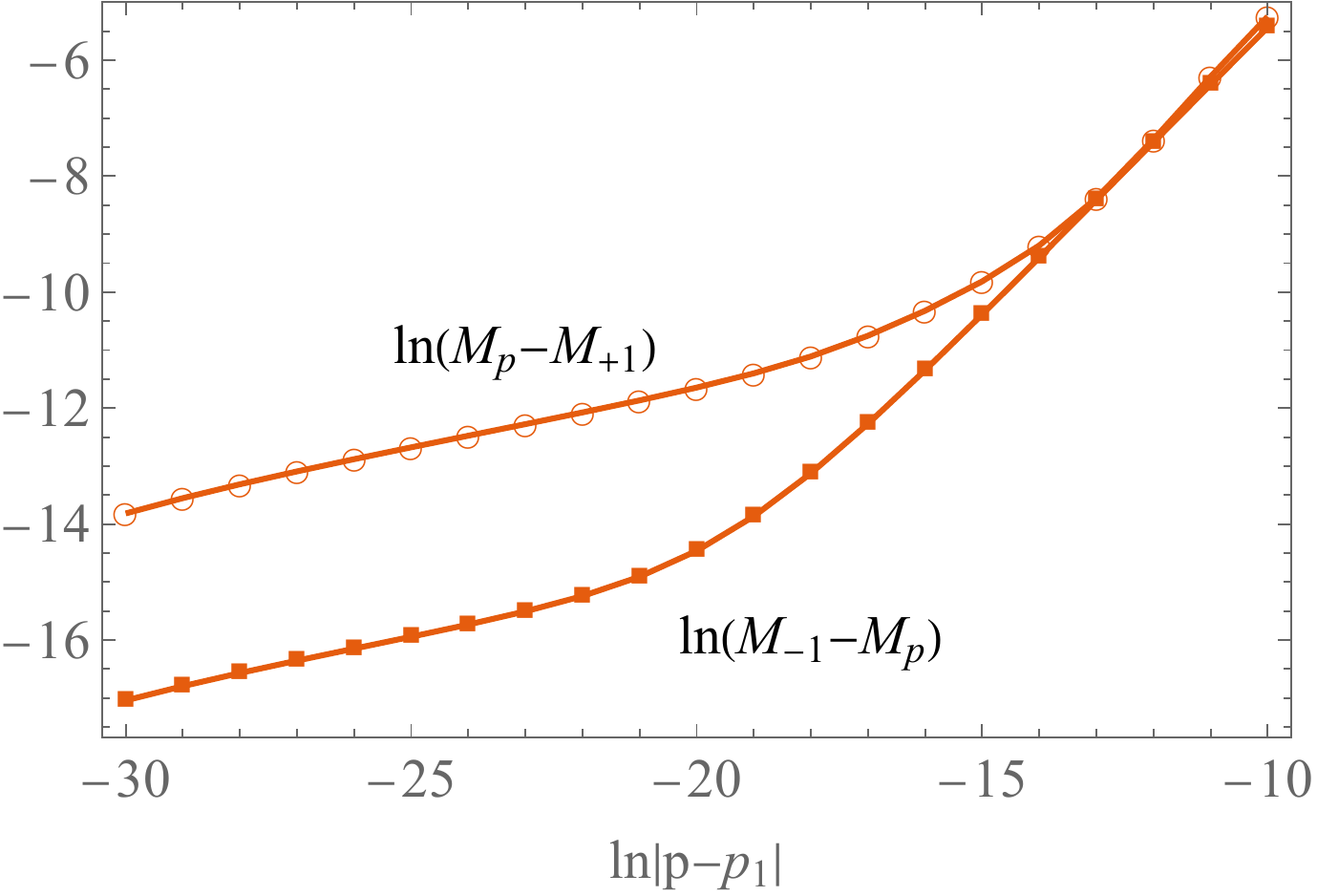} & \includegraphics[width=0.48\linewidth]{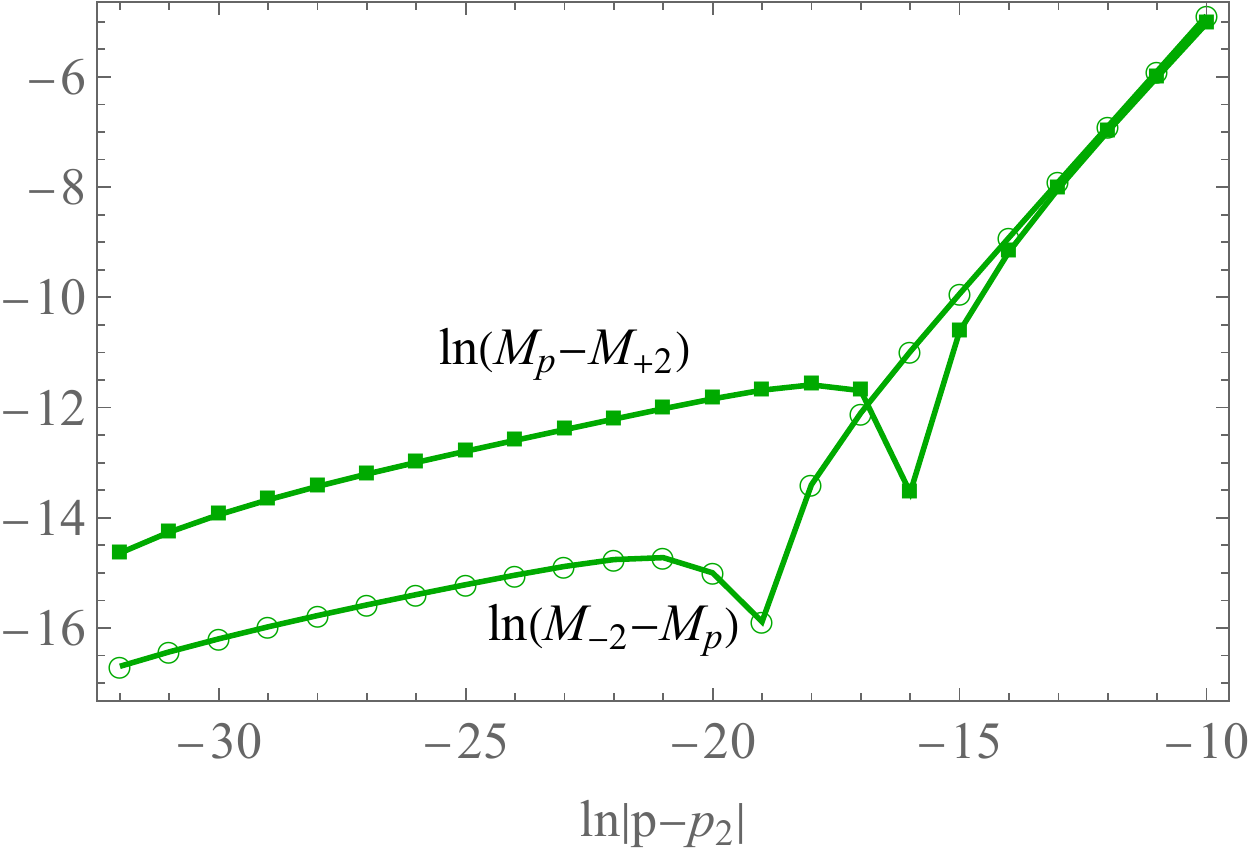}\tabularnewline
\end{tabular}
\par\end{centering}
{\footnotesize{}\caption{{\footnotesize{}\label{fig:RN0Mlnp}The power law between the irreducible
masses of the final black holes and the amplitude $p$ for initial
data (\ref{eq:initialRN}). The orange circles and squares are results
for subcritical and supercritical cases near $p_{1}$, respectively.
The green circles and squares are results for subcritical and supercritical
cases near $p_{2}$, respectively.}}
}{\footnotesize\par}
\end{figure}

Furthermore, in the late stages, when the system reaches equilibrium,
we observe the emergence of power laws that relate the irreducible
mass of the final black hole to the initial perturbation amplitude
$p$. Specifically, we find 
\begin{equation}
M_{-1}-M_{p}\propto(p_{1}-p){}^{\gamma_{-1}},\ \ \ M_{p}-M_{+1}\propto(p-p_{1})^{\gamma_{+1}},\label{eq:Mpsubp}
\end{equation}
for subcritical and supercritical cases, respectively. Here $M_{p}$
and $M_{\pm1}$ are the irreducible masses of the final black holes
resulting from the initial data (\ref{eq:initialRN}) with amplitude
$p$ and $p_{\pm1}\to p_{1}$ from above and below, respectively.
As depicted in the left panel of Fig.\ref{fig:RN0Mlnp}, for values
$\ln|p-p_{1}|\lesssim-20$, the indices $\gamma_{\pm1}\approx0.21$.
The values of $\gamma_{\pm1}$ are not universal and depend on the
families of initial data and system parameters, such as $M_{0},Q,\beta$
\cite{Zhang:2021nnn}. However, we consistently observe the power
laws (\ref{eq:Mpsubp}) in all cases of the critical dynamical transition.
These fractional indices are absent in type I critical gravitational
collapse. It is noteworthy that when $p$ deviates significantly from
$p_{1}$, the indices $\gamma_{\pm1}\approx1$. We propose that the
fractional power laws arise due to the matter escaping to spatial
infinity during the evolution. Indeed, we did not find fractional
indices in asymptotic AdS spacetime, where all matter remains confined,
resulting in indices always equal to 1 \cite{Zhang:2022cmu}. 

\begin{figure}[h]
\begin{centering}
\begin{tabular}{cc}
\includegraphics[width=0.48\linewidth]{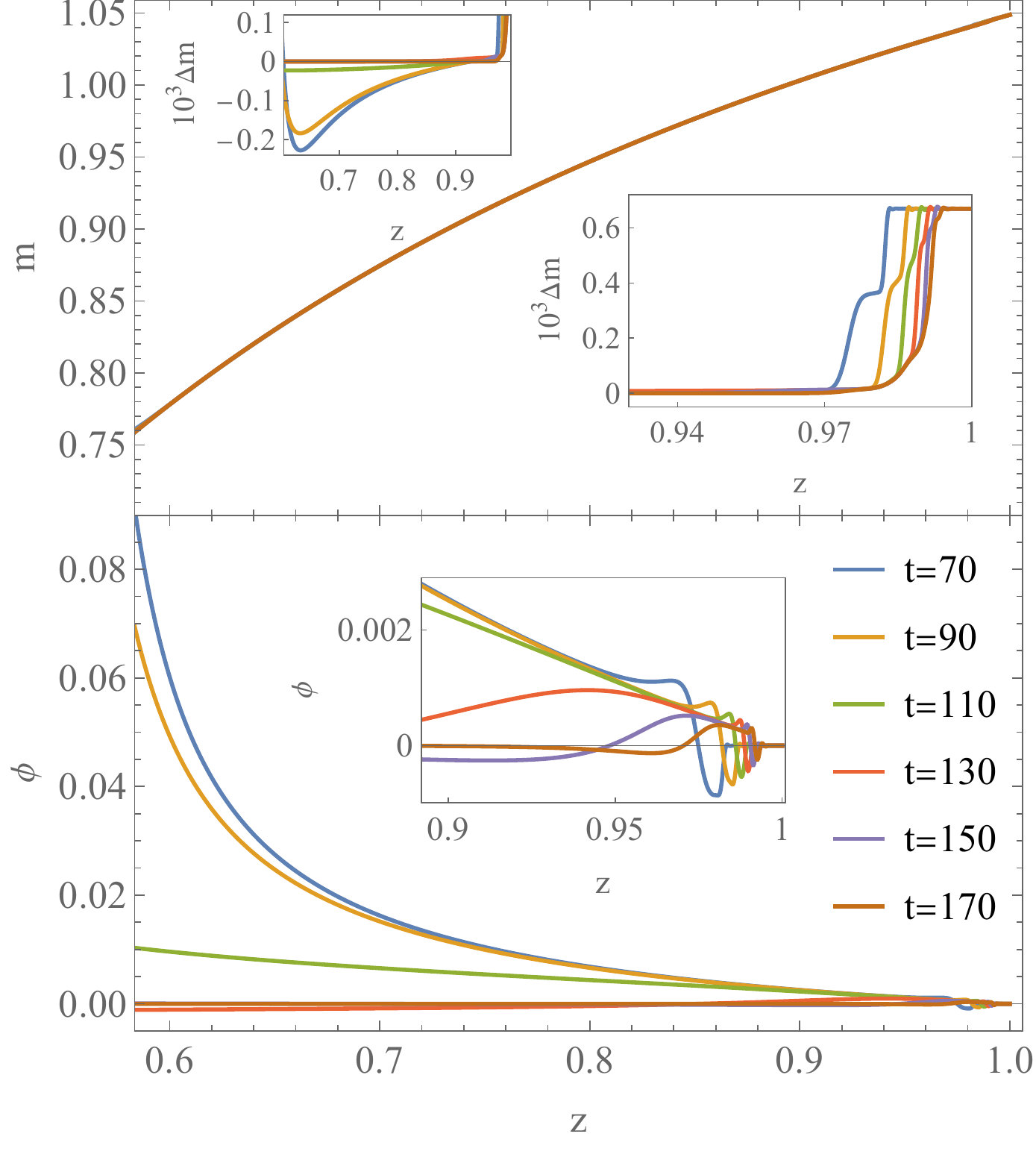} & \includegraphics[width=0.48\linewidth]{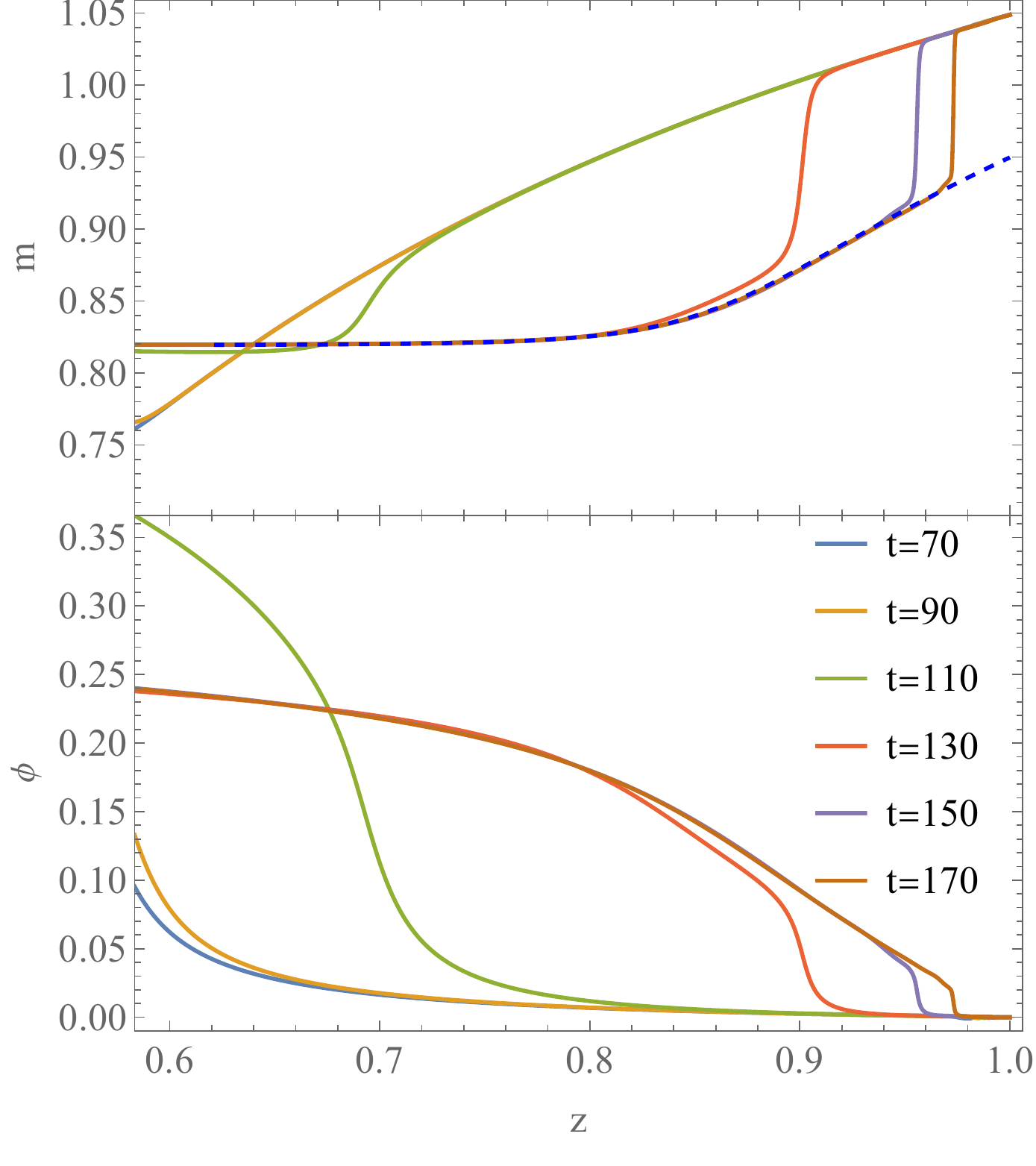}\tabularnewline
\end{tabular}
\par\end{centering}
{\footnotesize{}\caption{{\footnotesize{}\label{fig:RN0MSevo}The snapshots of the scalar field
and MS mass distribution at mid-late times. The left and right panels
depict the cases with initial amplitude $p=p_{1}-e^{-20}$ (subcritical)
and $p=p_{1}+e^{-20}$ (supercritical), respectively. The total mass
of the system is $M=1.04873$. The evolution of the scalar values
$\phi_{h}$ on the apparent horizon is illustrated in Fig.\ref{fig:RN0Mphihtp1}.
In the left panel, the insets present snapshots of $\Delta m=m_{RN}-m$,
where $m_{RN}$ is the MS mass of the final bald RN black hole with
a total mass of $M=1.04806$. In the right panel, the final SBH possesses
a total mass of $M=0.949646$, as indicated by the dashed blue curve. }}
}{\footnotesize\par}
\end{figure}

Finally, let us examine the evolution of the scalar field and the
MS mass distribution of the gravitational system at mid to late times.
This process corresponds to either the removal or the growth of the
scalar field of the intermediate CS for the subcritical or supercritical
cases, respectively. The snapshots of this evolution are depicted
in Fig.\ref{fig:RN0MSevo}. In the subcritical case, the scalar field
of the intermediate CS is absorbed by the central black hole, resulting
in the formation of a bald RN black hole. Only a small fraction of
energy (approximately $0.064\%$) escapes to spatial infinity. However,
the situation is quite different for the supercritical case. The scalar
field outside the black hole grows, leading to the formation of a
SBH. The energy of the scalar field comes from the Maxwell field.
At late times, around $9.45\%$ of the energy escapes to spatial infinity. 

\subsubsection{Dynamical critical behaviors of descalarization}

\begin{figure}[h]
\begin{centering}
\begin{tabular}{cc}
\includegraphics[width=0.48\linewidth]{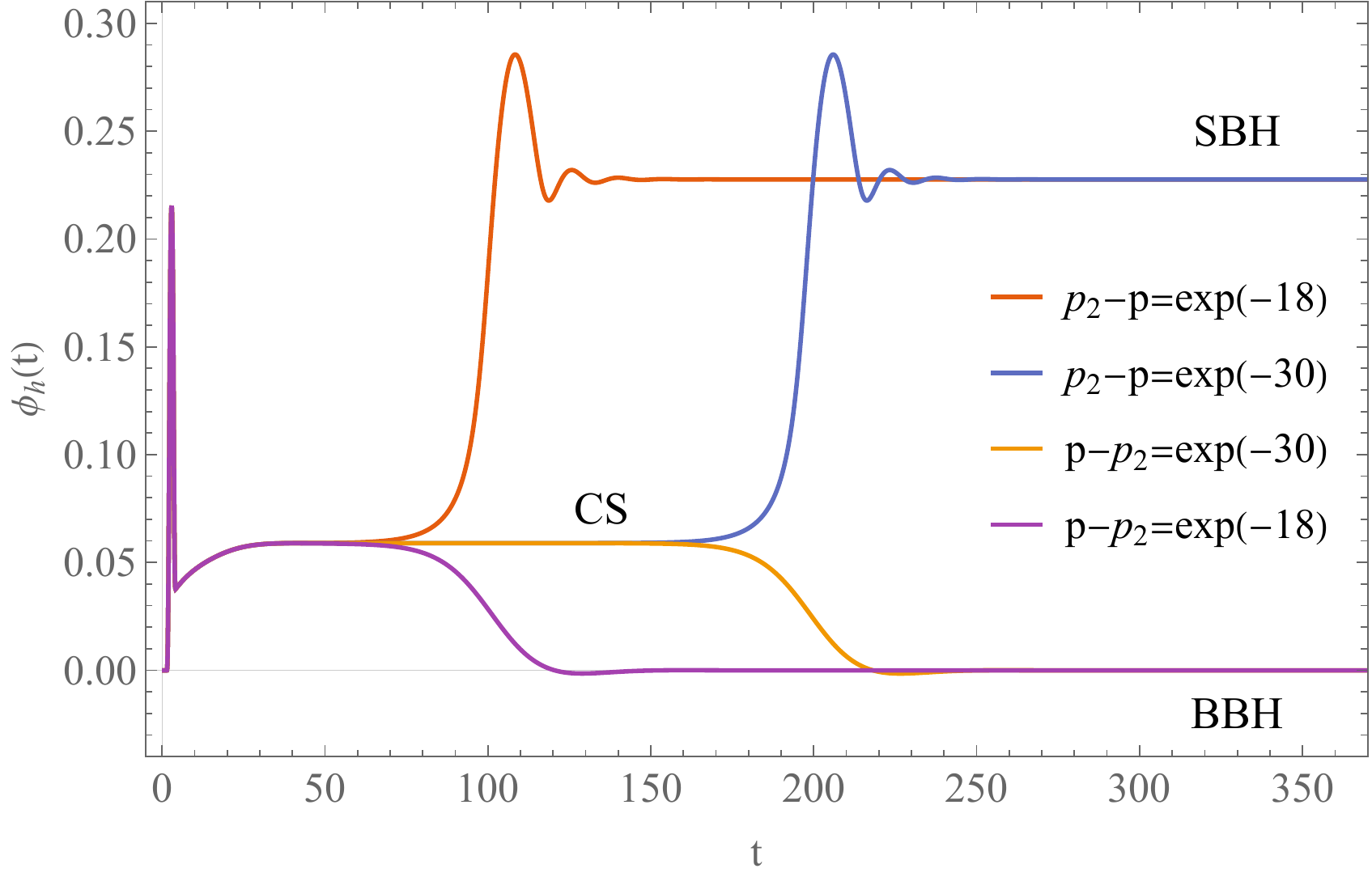} & \includegraphics[width=0.48\linewidth]{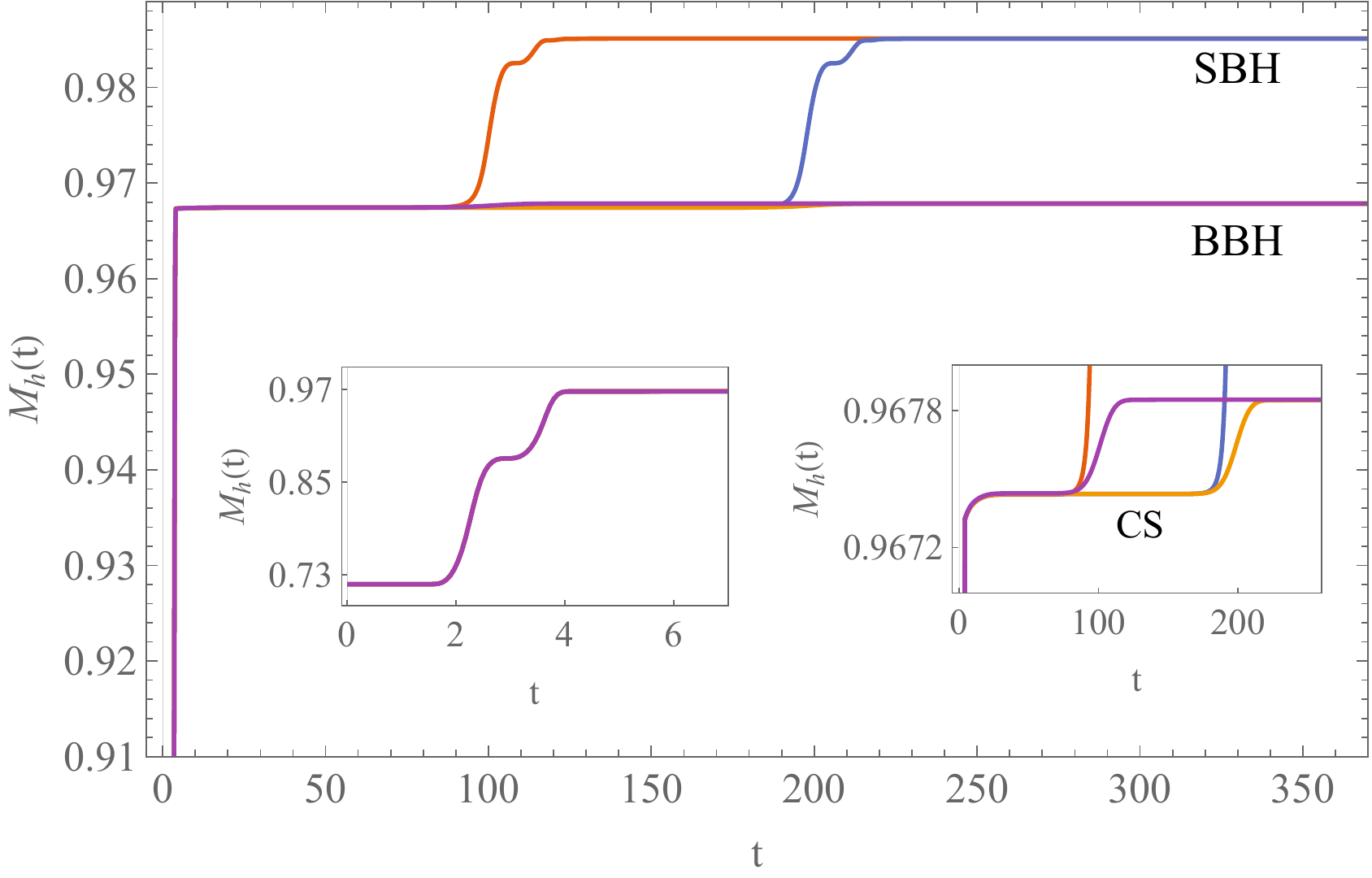}\tabularnewline
\end{tabular}
\par\end{centering}
{\footnotesize{}\caption{{\footnotesize{}\label{fig:RN0Mphihtp2}The evolution of the scalar
field value $\phi_{h}$ on the apparent horizon and the black hole
irreducible mass $M_{h}$ with respect to various $p$ near threshold
$p_{2}\approx0.2789936530457$ for initial data family (\ref{eq:initialRN}).
The insets in the right panel show the evolution of $M_{h}$ in the
early and middle times. The irreducible mass of the CS is very close
to that of the final BBH.}}
}{\footnotesize\par}
\end{figure}

Now we delve into the dynamics for descalarization near the threshold
$p_{2}$. As depicted in Fig.\ref{fig:RN0Mphihtp2}, we observe similar
dynamical critical behaviors. When $p<p_{2}$, the final outcome is
an SBH, whereas for $p>p_{2}$, it transforms into a bald RN black
hole. At the precise threshold $p_{2}$, a linearly unstable static
CS emerges. The closer $p$ is to $p_{2}$, the longer the intermediate
solutions remains on this CS. The scalar values $\phi_{h}$ on the
horizon of the intermediate CS and final SBH are almost the same with
those obtained for amplitudes near $p_{1}$. However, given the more
perturbation energy, here the irreducible mass $M_{h}$ of the intermediate
CS, final BBH and SBH are all significantly larger compared to those
obtained near $p_{1}$. It is worth noting that the final BBHs still
possess a smaller irreducible mass compared to the SBH near $p_{2}$.
In summary, the dynamical evolution process for $p$ near threshold
$p_{2}$ can be succinctly summarized as follows: 
\begin{equation}
\text{BBH (metastable) }+\text{perturbation}\to\begin{cases}
\text{SBH (metastable)}, & \text{subcritical }(p_{1}<p<p_{2}),\\
\text{CS (unstable SBH)}, & \text{critical }(p=p_{2}),\\
\text{bald BH (metastable)}, & \text{supercritical }(p>p_{2}).
\end{cases}\label{eq:p2CS}
\end{equation}

By analyzing the evolution of $\ln|\frac{d\phi_{h}}{dt}|$, we observe
similar patterns as depicted in Fig.\ref{fig:RN0phihdtlog}. The evolution
still can be divided into five distinct stages. Following a violent
change in the first stage, the intermediate solution gradually approaches
the CS in the second stage, exhibiting a damping rate of $\nu_{2}\approx0.14$.
In the third stage, the solution departs from the CS exponentially,
with an exponent of $\eta_{2}\approx0.123$. During the fourth stage,
the solution converges towards the final SBHin the subcritical case,
characterized by a dominant mode $\omega_{2s}\approx0.432-0.116i$.
Conversely, in the supercritical case, the solution evolves towards
the final BBH with a dominant mode $\omega_{2b}\approx0.098-0.090i$. 

The duration $T$ for which intermediate solution remains on the CS
still satisfies a similar relation in the form of (\ref{eq:Tlnp1}),
as shown in Fig.\ref{fig:RN0Tlnp}. The new coefficient $\gamma_{2}=\frac{1}{\eta_{2}}\approx8.14$.
Moreover, as shown in the right panel of Fig.\ref{fig:RN0Mlnp}, we
observe that the relations (\ref{eq:Mpsubp}) still hold. When $p$
is very close to $p_{2}$, and the corresponding coefficients $\gamma_{\pm2}\approx0.21$.
For $p$ that deviates significantly from $p_{2}$, the coefficients
$\gamma_{\pm2}\approx1$.

\subsection{Dynamical results when the seed black hole is a scalarized black
hole}

\begin{figure}[h]
\begin{centering}
\begin{tabular}{cc}
\includegraphics[width=0.48\linewidth]{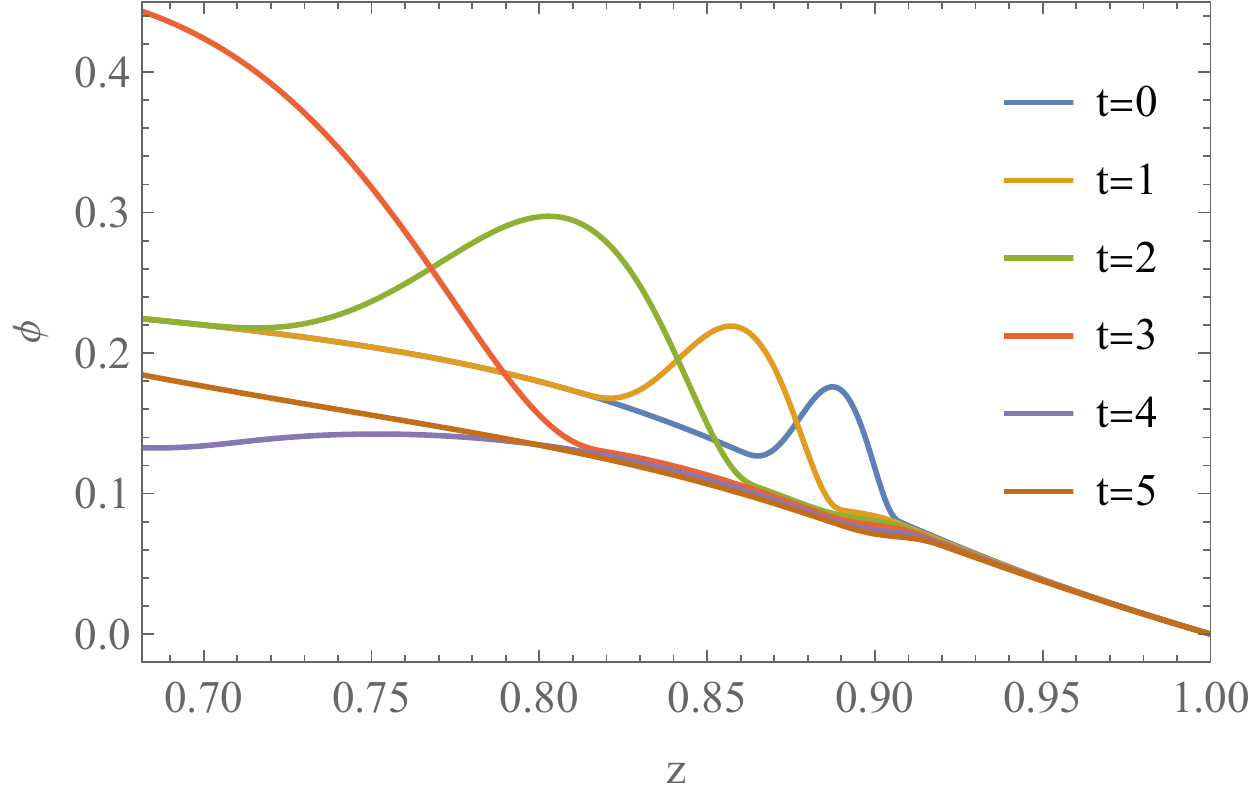} & \includegraphics[width=0.48\linewidth]{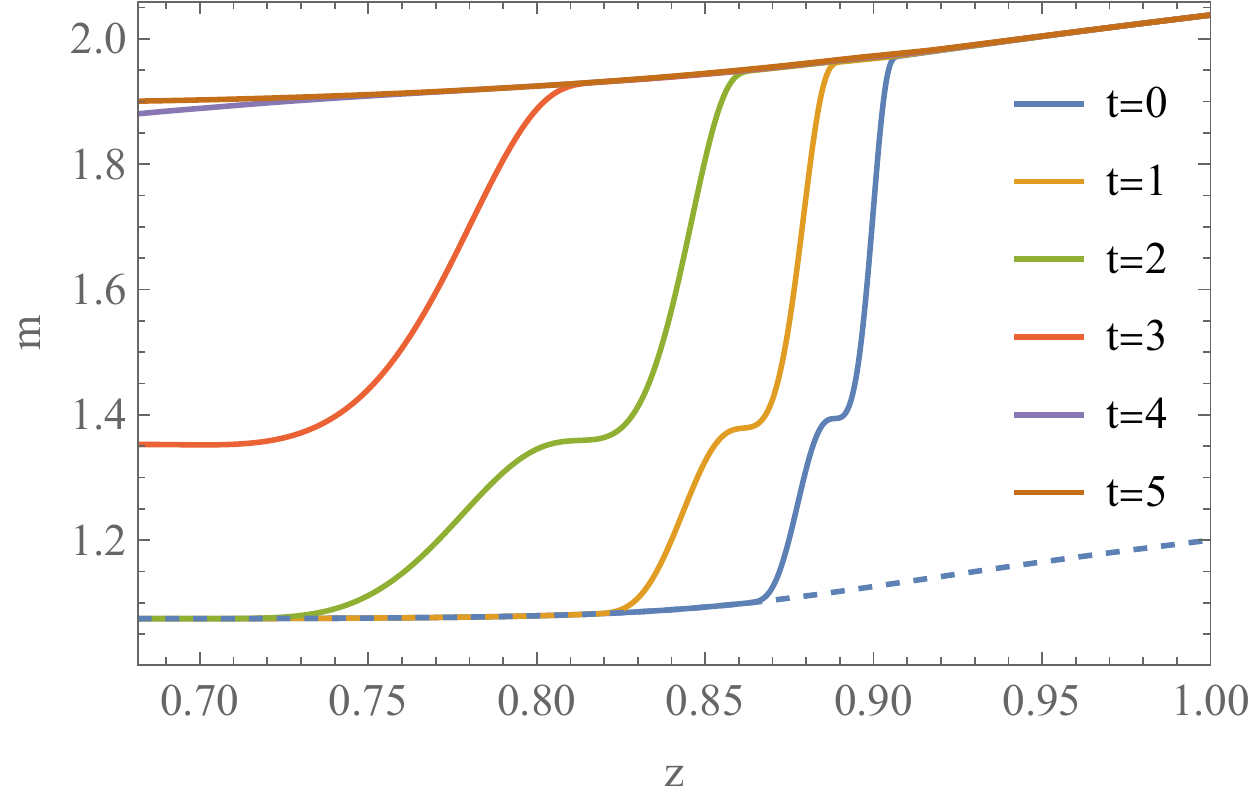}\tabularnewline
\end{tabular}
\par\end{centering}
{\footnotesize{}\caption{{\footnotesize{}\label{fig:SBH0MS}The early evolution of the scalar
field and MS mass when the seed black hole is an SBH with a total
mass $M=1.2$. In the right panel, the dashed blue curve represents
the MS mass for the seed SBH. The solid blue curve with $t=0$ represents
the MS mass of a nonequilibrium spacetime when the ingoing perturbation
(\ref{eq:initialSBH}) with amplitude $p=0.1285$ is applied. The
total mass increases to $M=2.038$. }}
}{\footnotesize\par}
\end{figure}

In this subsection, we investigate the cases when the seed black hole
takes the form of an SBH with a total mass of $M=1.2$. The coupling
parameter $\beta=2000$ and the black hole chagre $Q=0.9$ remain
fixed. The seed SBH exhibits a nontrivial distribution of the background
scalar field, characterized by a scalar charge of $Q_{s}=0.6939$.
To initiate the evolution, we apply an ingoing scalar field perturbation
using the expression (\ref{eq:initialSBH}). Fig.\ref{fig:SBH0MS}
presents the early evolution of the scalar field and the MS mass.
As depicted by the right panel, initially, the spacetime geometry
remains unchanged in the region $r<r_{1}$, resembling that of the
seed SBH. However, beyond the region $r>r_{2}$, the geometry undergoes
a transformation into a novel state that is distinct from both SBH
and RN black hole. The scalar perturbation propagates inward, driving
the evolution of the spacetime. 

\begin{figure}[h]
\begin{centering}
\begin{tabular}{cc}
\includegraphics[width=0.48\linewidth]{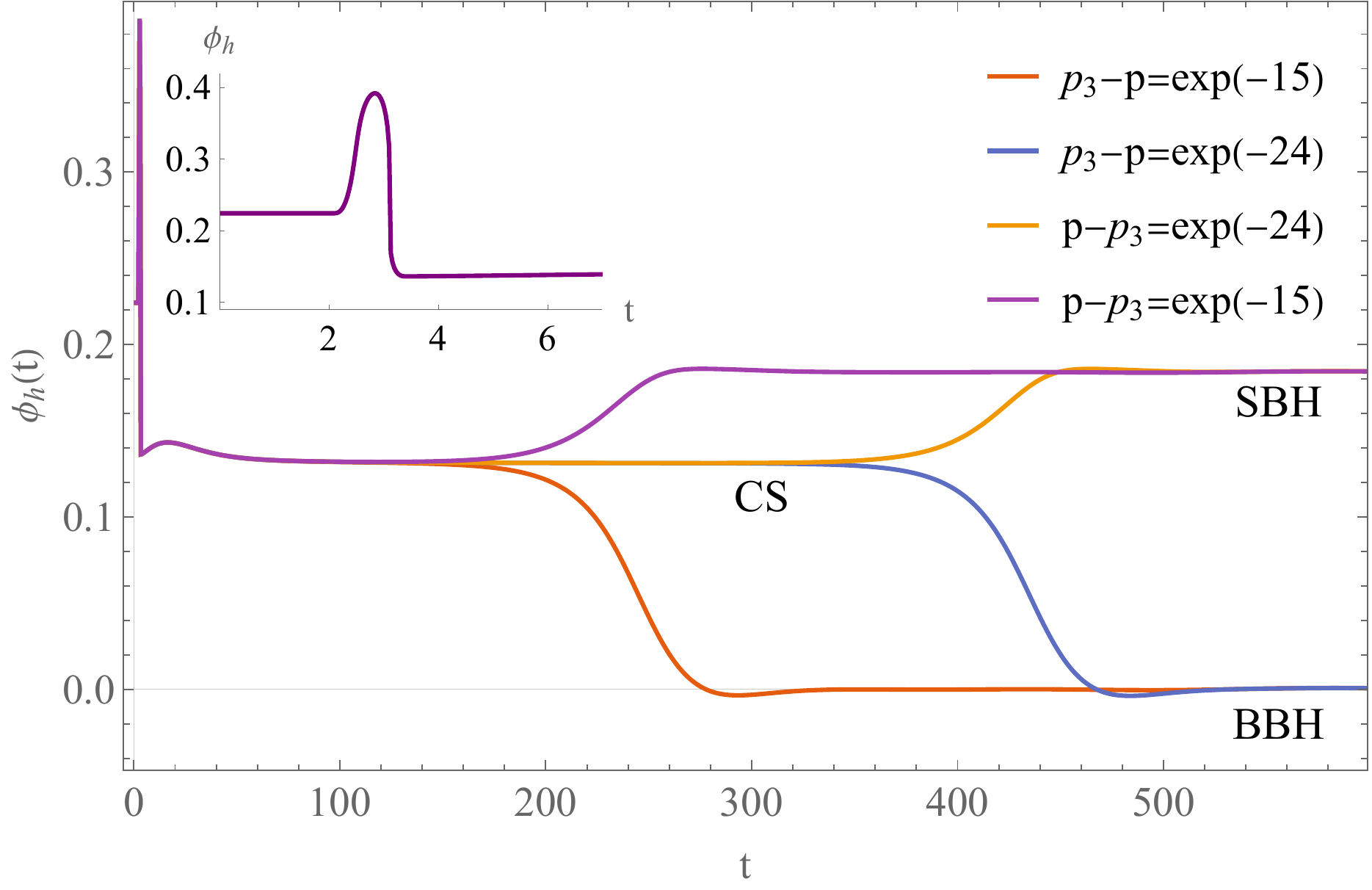} & \includegraphics[width=0.48\linewidth]{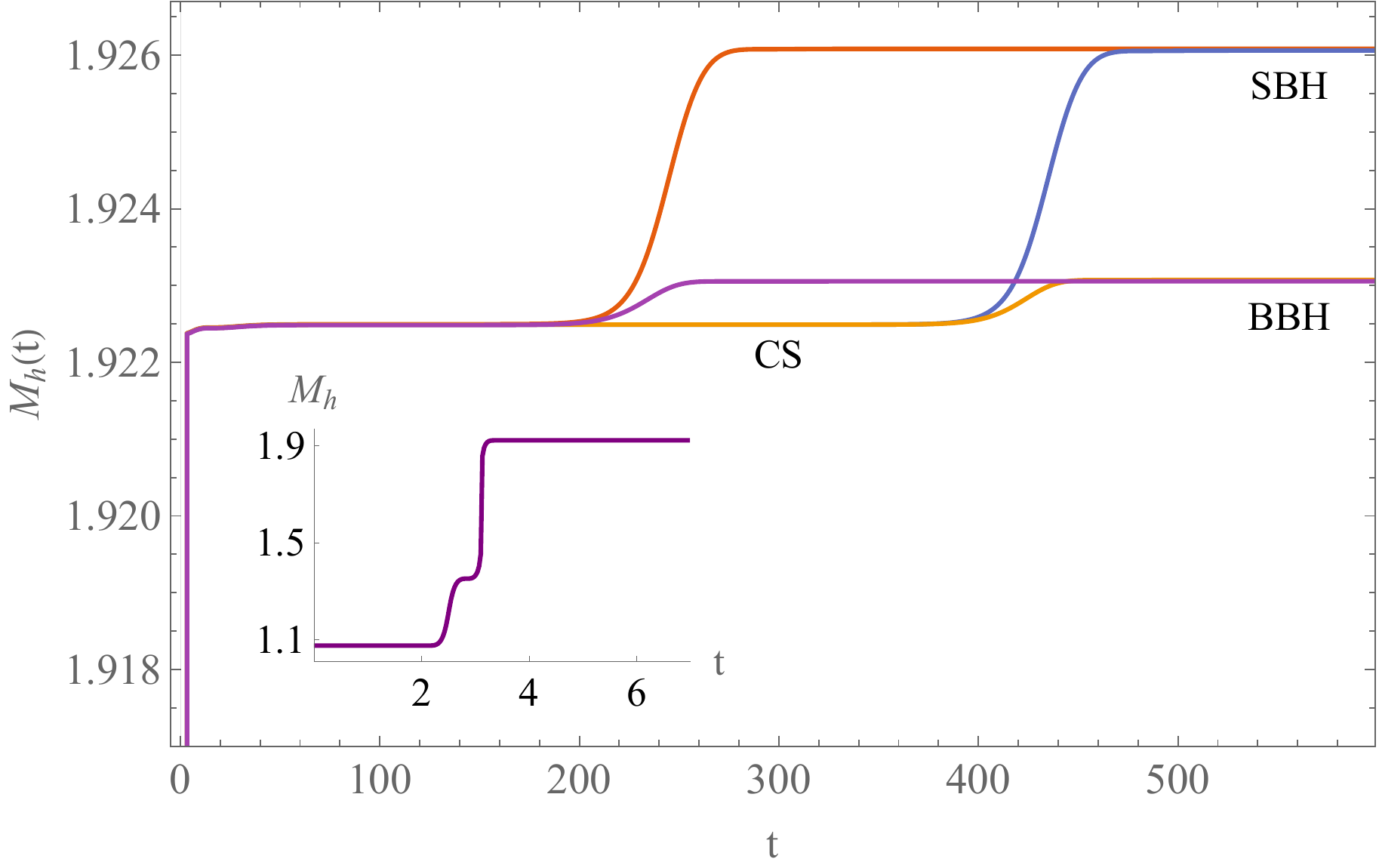}\tabularnewline
\end{tabular}
\par\end{centering}
{\footnotesize{}\caption{{\footnotesize{}\label{fig:SBH0Mphihtp3}The evolution of the scalar
field value $\phi_{h}$ on the apparent horizon and the black hole
irreducible mass $M_{h}$ with respect to various $p$ near threshold
$p_{3}\approx0.1284877802279$ for initial data family (\ref{eq:initialSBH}).
The insets show the evolution of $\phi_{h}$ and $M_{h}$ in the early
times. }}
}{\footnotesize\par}
\end{figure}

Unlike the cases described in subsection \ref{subsec:DynamicalRN},
for the initial data (\ref{eq:initialSBH}) parametrized by $p$,
we find only one threshold $p_{3}\approx0.1284877802279$. The final
black hole keeps scalarized when $p<p_{3}$. However, it undergoes
descalarization and becomes a bald RN black hole when $p>p_{3}$.
Near the threshold, we still observe type I critical dynamical behaviors.
The evolution of the scalar field value $\phi_{h}$ on the apparent
horizon and black hole irreducible mass $M_{h}$ when $p$ is close
to $p_{3}$ are depicted in Fig.\ref{fig:SBH0Mphihtp3}. Note that
the initia value of $\phi_{h}$ is nonzero here. When the ingoing
scalar perturbation reaches the black hole horizon, the system experiences
a drastic change. The scalar field increases fast and then drops fast.
The black hole irreducible mass increases a lot from $1.0745$ to
$1.9224$. Then the intermediate solution is attracted to a CS, and
remains on the CS for a duration satisfying $T\propto-\gamma_{3}\ln|p-p_{3}|$,
in which $\gamma_{3}=21.08$. At last, the intermediate solution departs
the CS exponentially with exponent $\eta_{3}=\frac{1}{\gamma_{3}}\approx0.0475$.
It converges to a final SBH when $p<p_{3}$ or to a final RN black
hole when $p>p_{3}$. 

We have found similar critical dynamical behaviors when the seed SBH
has a total mass of $M=1.8$ or $2$, where a linearly unstable CS
always emerges on the descalarization threshold. In summary, the critical
dynamics near $p_{3}$ for descalarization of an SBH can be expressed
as follow:
\begin{equation}
\text{SBH (metastable) }+\text{perturbation}\to\begin{cases}
\text{SBH (metastable)}, & \text{subcritical }(p<p_{3}),\\
\text{CS (unstable SBH)}, & \text{critical }(p=p_{3}),\\
\text{bald BH (metastable)}, & \text{supercritical }(p>p_{3}).
\end{cases}\label{eq:p3CS}
\end{equation}

\subsection{Comparison with type I critical gravitational collapse }

The critical dynamical scalarization and descalarization we have describled
above resemble the type I critical gravitational collapse, which can
be summarized as follows:
\begin{equation}
\text{flat space (metastable)}+\text{perturbation}\to\begin{cases}
\text{flat space (metastable)}, & \text{subcritical }(p<p_{\ast}),\\
\text{CS (unstable star)}, & \text{critical }(p=p_{\ast}),\\
\text{BH (stable)}, & \text{supercritical }(p>p_{\ast}).
\end{cases}
\end{equation}
Here the initial state is a flat spacetime. When the matter perturbation
is small, the matter will dissipates and leave a flat spacetime. But
when the perturbation is large, a black hole will be formed. At the
threshold, a CS emerges. The CS is a linearly unstable star. The duration
$T$ for which the intermediate solution remains on the CS still obey
a relation $T\propto-\gamma\ln|p-p_{*}|$. Here the coefficient $\gamma$
is a universal constant, regardless of the initial data family. This
is different from the critical dynamical scalarization and descalarization
here, in which the coefficient $\gamma$ is related to the initial
data family \cite{Zhang:2021nnn}. The source of this difference comes
from the fact that there is only one CS in the type I critical gravitational
collapse, but many CSs in the critical dynamical scalarization and
descalarization. This will be demonstrated in the next section. 

\section{Static solutions and quasinormal modes \label{sec:Static}}

In the previous section, we discovered the crucial role of CSs in
the critical dynamics near the threshold. These CSs are, in fact,
statically linearly unstable SBHs, while the final SBH and BBH are
static linearly stable solutions. In this section, we solve the static
equations of motion directly, and delve into the thermodynamic property
and perturbative stability of the CS, final SBH and BBH. 

\subsection{Equations of motion for static solutions and their perturbations
\label{subsec:Static-solutions}}

To investigate the static solutions and their perturbative properties
in a spherical context, we adopt the following metric ansatz: 
\begin{equation}
ds^{2}=-\tilde{N}e^{-2\tilde{\delta}}dt_{s}^{2}+\frac{1}{\tilde{N}}dr^{2}+r^{2}(d\theta^{2}+\sin^{2}\theta d\phi^{2}).\label{eq:Sch}
\end{equation}
Here, the metric functions $\tilde{N}$ and $\tilde{\delta}$ depend
on $t,r$ and take the following forms: 
\begin{equation}
\tilde{N}(t,r)=N(r)+\epsilon N_{1}(r)e^{-i\omega t},\ \ \ \tilde{\delta}(t,r)=\delta(r)+\epsilon\delta_{1}(r)e^{-i\omega t}.
\end{equation}
The scalar field is assumed to be
\begin{equation}
\tilde{\phi}(t,r)=\phi(r)+\epsilon\phi_{1}(r)e^{-i\omega t}.\label{eq:wt}
\end{equation}
The Maxwell field is determined by $\partial_{r}A=\frac{Qe^{-\tilde{\delta}}}{r^{2}f(\tilde{\phi})}$
where $Q$ represents the electric charge of the black hole. Here,
$N,\delta$, and $\phi$ are background metric and scalar functions
for the static solutions, $\epsilon$ is the control parameter in
the linear expansion, and the complex quantity $\omega=\omega_{R}+i\omega_{I}$
corresponds to the quasinormal modes or eigenvalues of the perturbative
eigenstates $N_{1},\delta_{1}$, and $\phi_{1}$. By substituting
these ansatz into the equations of motion for gravity and scalar fields,
and expanding the equations with respect to $\epsilon$, we obtain
the following leading-order equations.
\begin{align}
\partial_{r}N & =\frac{1-N}{r}-r(\partial_{r}\phi)^{2}N-\frac{Q^{2}}{r^{3}f(\text{\ensuremath{\phi}})},\nonumber \\
\text{\ensuremath{\partial_{r}\delta}} & =-r(\partial_{r}\phi)^{2},\label{eq:ddr}\\
\partial_{r}^{2}\phi & =\frac{1}{rN}\left(N+1-\frac{Q^{2}}{r^{2}f(\phi)}\right)\partial_{r}\phi+\frac{Q^{2}}{2r^{4}Nf^{2}(\phi)}\frac{df(\phi)}{d\phi}.\nonumber 
\end{align}
At the subleading order, we find that the metric perturbations can
be expressed in terms of scalar field perturbation as
\begin{align}
N_{1}= & -2rN\phi_{1}\partial_{r}\phi,\ \ \ \partial_{r}\delta_{1}=-2r\partial_{r}\phi\partial_{r}\phi_{1}.
\end{align}
Then the equation for scalar field perturbation is decoupled from
the metric perturbations and can be reduced to a Schrödinger-like
equation: 
\begin{equation}
\left(\frac{d^{2}}{dr_{*}^{2}}+\omega^{2}-V_{\text{eff}}\right)\psi=0.\label{eq:scalarPert}
\end{equation}
Here we have introduced the tortoise coordinate $r_{*}$ by $\frac{dr_{*}}{dr}=\frac{1}{Ne^{-\delta}}$,
and defined $\psi(r)=r\phi_{1}(r)$. The effective potential \cite{Fernandes:2019rez}
\begin{align}
V_{\text{eff}}= & \frac{Ne^{-2\delta}}{r^{2}}\left[1-N-2r^{2}\phi'{}^{2}-\frac{Q^{2}}{r^{2}f}\left(1-2r^{2}\phi'{}^{2}+\frac{2r\phi'\dot{f}}{f}+\frac{f\ddot{f}-2\dot{f}{}^{2}}{2f^{2}}\right)\right],\label{eq:Veff}
\end{align}
where $\phi'=\partial_{r}\phi,$ $\dot{f}=\frac{df(\phi)}{d\phi}$,
and $\ddot{f}=\frac{d^{2}f(\phi)}{d\phi^{2}}$. 

We can solve the static background metric and scalar functions using
equations (\ref{eq:ddr}), and then substitute them into equations
(\ref{eq:scalarPert},\ref{eq:Veff}) to determine the quasinormal
mode $\omega$ under appropriate boundary conditions. However, in
order to directly compare the results obtained from the dynamical
approach, we perform a coordinate transformation
\begin{equation}
dt_{s}=dt-\frac{\zeta}{(1-\zeta^{2})\alpha}dr,
\end{equation}
to convert to the PG coordinates (\ref{eq:PG}). This transformation
is feasible when the background are time-independent. The metric functions
are related as 
\begin{equation}
N=1-\zeta^{2},\ \ \ e^{-\delta}=\alpha.\label{eq:trans}
\end{equation}
For static background solutions, the equations of motion (\ref{eq:ddr})
turn to be 
\begin{align}
0= & \partial_{r}\alpha-r\alpha(\partial_{r}\phi)^{2},\label{eq:sar}\\
0= & \partial_{r}\zeta+\frac{\zeta}{2r}-\frac{r(1-\zeta^{2})}{2\zeta}(\partial_{r}\phi)^{2}-\frac{Q^{2}}{2r^{3}\zeta f(\phi)},\label{eq:szr}\\
0= & \partial_{r}^{2}\phi+\frac{1}{\left(\zeta{}^{2}-1\right)}\left[\left(\frac{Q^{2}}{r^{2}f(\phi)}+\zeta{}^{2}-2\right)\frac{\partial_{r}\phi}{r}-\frac{Q^{2}}{2r^{4}f^{2}(\phi)}\frac{df(\phi)}{d\phi}\right].\label{eq:sprr0}
\end{align}
The scalar perturbation equation (\ref{eq:scalarPert}) can be rewritten
as 
\begin{align}
\left[(1-\zeta^{2})\alpha\frac{\partial}{\partial r}\left((1-\zeta^{2})\alpha\frac{\partial}{\partial r}\right)+\omega^{2}-V_{\text{eff}}\right]\psi & =0,\label{eq:sprr}
\end{align}
where the variables $N,\delta$ in effective potential $V_{\text{eff}}$
should be replaced by (\ref{eq:trans}). 

\subsection{Static solutions}

In this subsection, our focus is on finding the solutions for the
static background. It is worth noting that the equations (\ref{eq:szr},\ref{eq:sprr0})
are decoupled from (\ref{eq:sar}). This decoupling allows us to initially
solve for $\zeta,\phi$ using (\ref{eq:szr},\ref{eq:sprr0}) at first,
and subsequently determine $\alpha$ using (\ref{eq:sar}). To solve
the static equations (\ref{eq:szr},\ref{eq:sprr0}), it is essential
to specify appropriate boundary conditions.

\subsubsection{Boundary conditions and numerical setup}

At spatial infinity, the solutions can be expanded as 
\begin{align}
\zeta= & \sqrt{\frac{2M}{r}}\left(1-\frac{Q^{2}+Q_{s}^{2}}{4Mr}+\cdots\right),\label{eq:zpinf}\\
\phi= & \frac{Q_{s}}{r}+\frac{MQ_{s}}{r^{2}}+\cdots.\nonumber 
\end{align}
Here $M$ denotes the total mass of the system, and $Q,Q_{s}$ denotes
the electric and scalar charge, respectively. Near the event horizon
$r_{H}$ of the black hole, the solutions can be expanded as 
\begin{align}
\zeta= & 1+\frac{1}{2r_{H}}\left(\frac{Q^{2}}{f(\phi_{H})r_{H}^{2}}-1\right)(r-r_{H})+\cdots,\label{eq:bdz}\\
\phi= & \phi_{H}+\frac{Q^{2}}{2r_{H}f(\phi_{H})\left[Q^{2}-r_{H}^{2}f(\phi_{H})\right]}\frac{df(\phi_{H})}{d\phi}(r-r_{H})+\cdots,\label{eq:bdp}
\end{align}
where $\phi_{H}$ is the scalar value on the event horizon. Note that
in EMS theory, the scalar hair is a secondary hair \cite{Herdeiro:2015waa}.
Given $\beta,Q$ and $M$, the scalar charge $Q_{s}$ or $\phi_{H}$
is determined.

By providing values for $\beta,Q,r_{H}$, and an initial guess value
for $\phi_{H}$, and imposing the boundary conditions (\ref{eq:bdz},\ref{eq:bdp})
at $r_{b1}=r_{H}(1+\epsilon)$, typically with $\epsilon\approx10^{-7}$,
we can numerically integrate the equations (\ref{eq:szr},\ref{eq:sprr0})
up to $r_{b2}$ which is typically around $10^{6}r_{H}$. A static
background solution is found if the solution $\phi(r)$ approaches
zero at $\text{\ensuremath{r_{b2}}}$. If the solution $\phi(r)$
does not approach zero at $r_{b2}$, we need to adjust the initial
guess value $\phi_{H}$ and repeat the integration process. Once we
obtain static background solutions $\zeta,\phi$, we can use equation
(\ref{eq:zpinf}) to calculate the total mass $M$ and scalar charge
$Q_{s}$, given by 
\begin{equation}
M=\lim_{r\to\infty}\frac{r}{2}\zeta(r)^{2},\ \ \ Q_{s}=-\lim_{r\to\infty}r^{2}\partial_{r}\phi(r).
\end{equation}

At spatial infinity, $\alpha$ can be expanded as 
\begin{align}
\alpha & =\alpha_{0}\left(1-\frac{Q_{s}^{2}}{2r^{2}}+\cdots\right),\label{eq:ainf}
\end{align}
where $\alpha_{0}$ is an undetermined constant due to the auxiliary
freedom in $\alpha dt$ in PG coordinate. In order to be consistent
with the dynamical method, we set $\alpha_{0}=1$. The equation (\ref{eq:sar})
can be solved by using above boundary expansion at $r_{b2}$. Now
we can determine the black hole temperature 
\begin{equation}
T_{H}=\frac{N'(r_{H})}{4\pi}e^{-\delta(r_{H})}=\frac{\zeta'(r_{H})}{2\pi}\alpha(r_{H})=\frac{\alpha(r_{H})}{4\pi r_{H}}\left(1-\frac{Q^{2}}{f(\phi_{H})r_{H}^{2}}\right).\label{eq:Temperature}
\end{equation}
The temperature of a RN black hole is given by $T_{H}=\frac{1}{4\pi r_{H}}\left(1-\frac{Q^{2}}{r_{H}^{2}}\right).$
To facilitate our analysis, we will utilize the reduced temperature
$t_{H}=8\pi MT_{H}$ and the reduced entropy $s_{H}=\frac{S}{4\pi M^{2}}$,
where the black hole entropy is denoted by $S=\frac{A_{H}}{4}=\pi r_{H}^{2}.$ 

\begin{figure}[h]
\begin{centering}
\begin{tabular}{cc}
\includegraphics[width=0.48\linewidth]{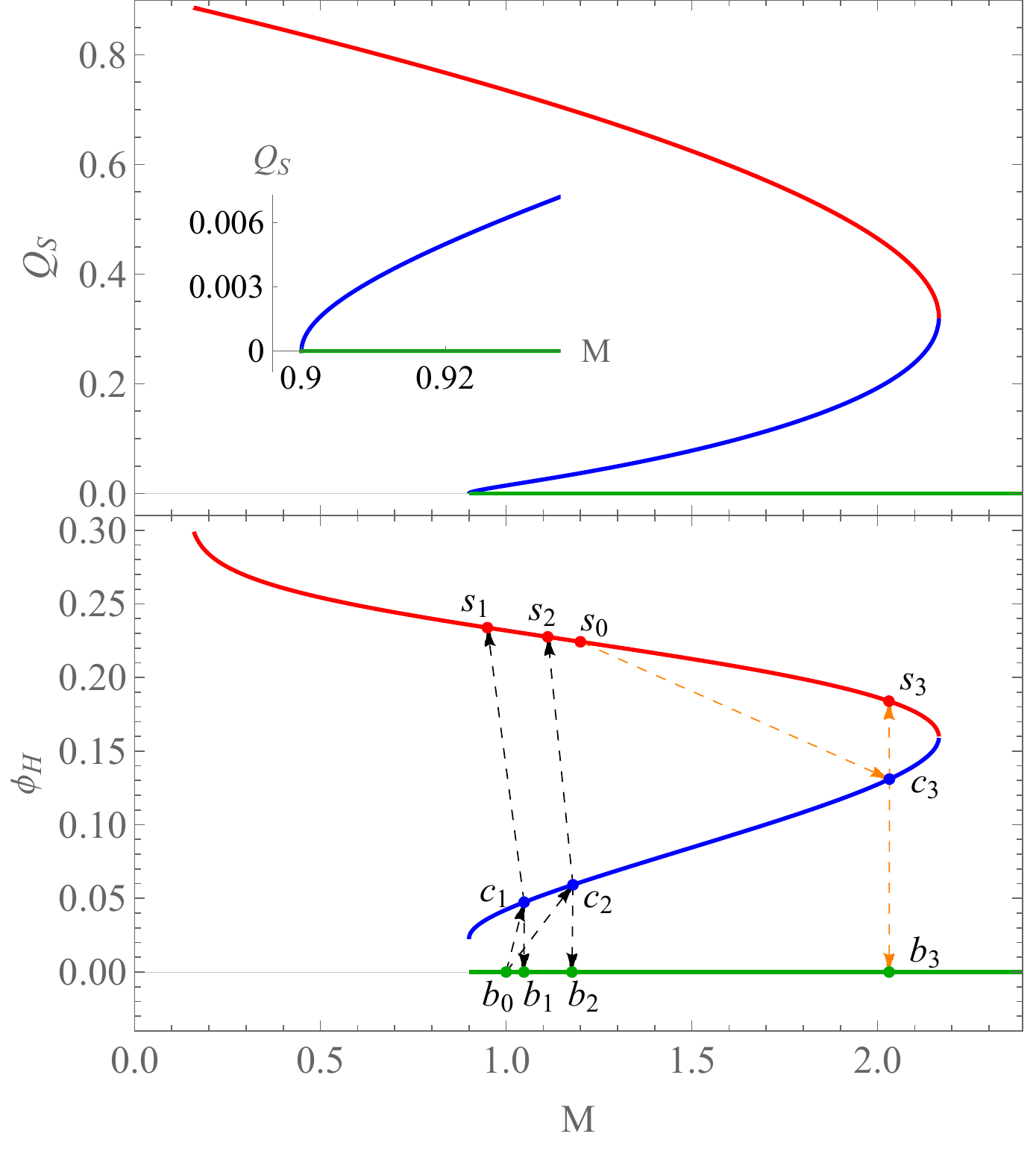} & \includegraphics[width=0.48\linewidth]{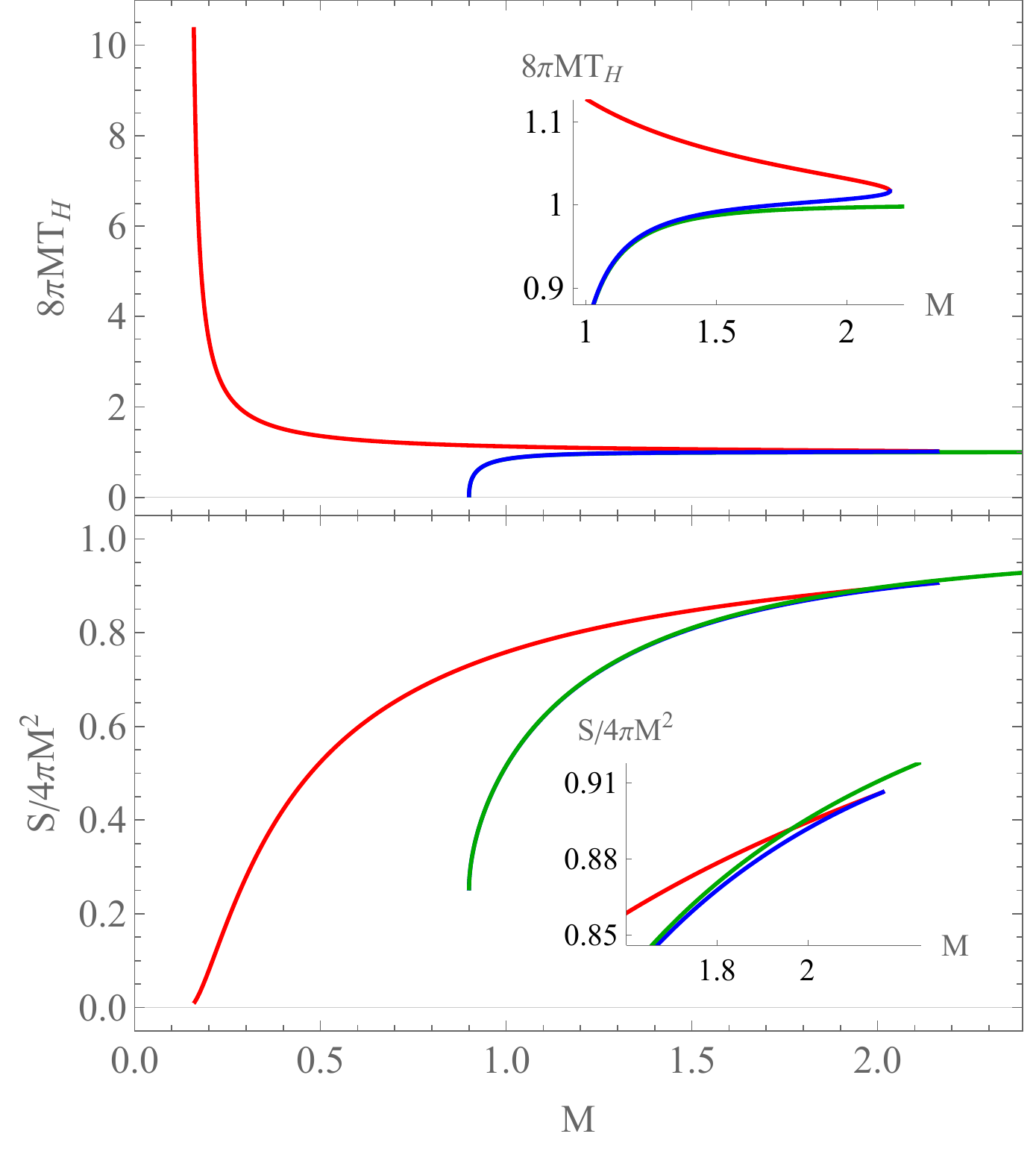}\tabularnewline
\end{tabular}
\par\end{centering}
{\footnotesize{}\caption{{\footnotesize{}\label{fig:static}The scalar charge $Q_{s}$, scalar
value $\phi_{H}$ on the event horizon, reduced temperature $t_{H}$,
and reduced entropy $s_{H}$ are illustrated for static solutions
with $\beta=2000$ and $Q=0.9$. At the same total mass $M$, we observe
two branches of SBHs. The blue and red curves correspond to SBHs with
lower and higher temperatures, respectively. The green curve represents
the BBH. The cold SBHs exhibit temperatures and entropies that closely
resemble those of the BBH. The differences between them are highlighted
in the insets.}}
}{\footnotesize\par}
\end{figure}

\subsubsection{The results for static solutions}

In Fig.\ref{fig:static}, we present various properties of static
solutions, including the scalar charge $Q_{s}$, scalar value $\phi_{H}$
on the event horizon, the reduced temperature $t_{H}$, and reduced
entropy $s_{H}$. These solutions are obtained with $\beta=2000$
and $Q=0.9$. We observe the presence of a bald RN black hole whenever
the total mass $M$ exceeds the charge $Q=0.9$. Additionally, we
identify two distinct branches of SBH solutions. The blue branch corresponds
to cold SBHs characterized by lower temperatures, while the red branch
represents hot SBH with higher temperatures. These branches intersect
at $(M,Q_{s},\phi_{H})=(2.165,0.3214,0.1594)$. It appears that the
cold branch bifurcates from the extremal RN black hole. However, the
situation near the extremal black hole is highly intricate, and for
a more comprehensive understanding, we refer readers to \cite{Blazquez-Salcedo:2020crd}.
Within the range of $M\in(0.9,2.165)$, the cold SBHs, hot SBHs, and
BBHs coexist. In this region, the cold SBHs always possess the smallest
entropy, making them thermodynamically unfavorable. As for the hot
SBHs, their entropy are either greater or smaller than those of the
BBH, depending on whether $M$ lies in the range $(0.9,1.964)$ or
$(1.964,2.165)$.

\subsubsection{Comparison with dynamical results}

When comparing with the dynamical outcomes, we observe that the critical
dynamical scalarization and descalarization near the thresholds, initiated
from the initial perturbation (\ref{eq:initialRN}) applied to the
seed bald RN black hole, correspond to the following trajectories:
\begin{equation}
b_{0}\to c_{1}\to\begin{cases}
b_{1}, & (p<p_{1}),\\
c_{1}, & (p=p_{1}),\\
s_{1}, & (p_{1}<p<p_{2}),
\end{cases}\ \ \ b_{0}\to c_{2}\to\begin{cases}
s_{2}, & (p_{1}<p<p_{2}),\\
c_{2}, & (p=p_{2}),\\
b_{2}, & (p>p_{2}).
\end{cases}
\end{equation}
The critical dynamical descalarization, initiated from the seed SBH
through the initial perturbation (\ref{eq:initialSBH}), follows the
following trajectory:
\begin{equation}
s_{0}\to c_{3}\to\begin{cases}
s_{3}, & (p<p_{3}),\\
c_{3}, & (p=p_{3}),\\
b_{3}, & (p>p_{3}).
\end{cases}
\end{equation}
The characteristics of these static solutions are presented in table
\ref{tab:DynamicsBCH}. Concerning the final equilibrium states, both
scalarization and descalarization can be interpreted as first-order
phase transitions, given that all the quantities exhibit a gap in
the vicinity of the threshold.

\begin{table}[h]
\begin{centering}
\begin{tabular}{|c|cccc||ccc||cccc|}
\hline 
 & $b_{0}$ & $c_{1}$ & $b_{1}$ & $s_{1}$ & $c_{2}$ & $s_{2}$ & $b_{2}$ & $s_{0}$ & $c_{3}$ & $s_{3}$ & $b_{3}$\tabularnewline
\hline 
$M$ & {\footnotesize{}1} & {\footnotesize{}1.0485} & {\footnotesize{}1.0480} & {\footnotesize{}0.9496} & {\footnotesize{}1.1799} & {\footnotesize{}1.1121} & {\footnotesize{}1.1772} & {\footnotesize{}1.2} & {\footnotesize{}2.0320} & {\footnotesize{}2.0305} & {\footnotesize{}2.0311}\tabularnewline
\hline 
$Q_{s}$ & {\footnotesize{}0} & {\footnotesize{}0.0200} & {\footnotesize{}0} & {\footnotesize{}0.7454} & {\footnotesize{}0.0348} & {\footnotesize{}0.7125} & {\footnotesize{}0} & {\footnotesize{}0.6938} & {\footnotesize{}0.2050} & {\footnotesize{}0.4503} & {\footnotesize{}0}\tabularnewline
\hline 
$\phi_{H}$ & {\footnotesize{}0} & {\footnotesize{}0.0474} & {\footnotesize{}0} & {\footnotesize{}0.2338} & {\footnotesize{}0.0592} & {\footnotesize{}0.2276} & {\footnotesize{}0} & {\footnotesize{}0.2242} & {\footnotesize{}0.1310} & {\footnotesize{}0.1839} & {\footnotesize{}0}\tabularnewline
\hline 
$t_{H}$ & {\footnotesize{}0.5851} & {\footnotesize{}0.8971} & {\footnotesize{}0.6882} & {\footnotesize{}1.1377} & {\footnotesize{}0.9553} & {\footnotesize{}1.1080} & {\footnotesize{}0.8969} & {\footnotesize{}1.0956} & {\footnotesize{}1.0079} & {\footnotesize{}1.0296} & {\footnotesize{}1.7197}\tabularnewline
\hline 
$s_{H}$ & {\footnotesize{}0.5154} & {\footnotesize{}0.5713} & {\footnotesize{}0.5718} & {\footnotesize{}0.7447} & {\footnotesize{}0.6757} & {\footnotesize{}0.7845} & {\footnotesize{}0.6762} & {\footnotesize{}0.8019} & {\footnotesize{}0.8952} & {\footnotesize{}0.8969} & {\footnotesize{}0.8992}\tabularnewline
\hline 
\end{tabular}
\par\end{centering}
\caption{{\footnotesize{}\label{tab:DynamicsBCH}The total mass $M$, scalar
charge $Q_{s}$, scalar value $\phi_{H}$ on the event horizon, reduced
temperature $t_{H}$, and reduced entropy $s_{H}$ for the bald RN
black holes $b_{0,1,2}$, critical solutions or cold SBHs $c_{1,2,3}$,
and hot SBHs $s_{0,1,2,3}$. }}
\end{table}

From table \ref{tab:DynamicsBCH}, we observe that the intermediate
CS consistently possesses a larger total mass than the corresponding
final SBH or BBH, as a result of energy escaping to spatial infinity.
Additionally, the intermediate CS exhibits a smaller entropy compared
to the corresponding final SBH or BBH, in accordance with the second
law of black hole thermodynamics. Interestingly, the dynamics do not
precisely align with thermodynamic expectations. For instance, the
bald RN black hole has a smaller entropy than the hot SBH when $M\in(0.9,1.964)$.
However, when the seed black hole is an RN black hole with $M=1$,
subjected to perturbation (\ref{eq:initialRN}) with amplitude $p_{1}<p<p_{2}$,
the final equilibrium states of the dynamic evolution are the hot
SBH with smaller entropy, rather than the RN black hole with larger
entropy, as depicted in the lower left panel of Fig.\ref{fig:static}. 

We compare our findings with those obtained in a generalized scalar-tensor
theory, as presented in a previous study \cite{Liu:2022fxy}. In that
work, the dynamical scalarization and descalarization were investigated
within a framework where the scalar field couples to both the Gauss-Bonnet
curvature and the Ricci curvature. We observe that the critical dynamical
scalarization exhibits typically type I behaviors to what we have
found here. However, the critical dynamical descalarization displays
intriguing differences. In study \cite{Liu:2022fxy}, the descalarization
manifests as a first-order phase transition, from a thermodynamic
standpoint. However, the phase transition point occurs at the junction
where the hot and cold branches of  SBH merge. In contrast, within
the EMS theory presented here, the phase transition point always precedes
the junction point. It will be worth exploring whether there are deeper
physical reasons for this difference. 

\subsection{Perturbation and QNM}

Now we turn to solve the perturbative equation (\ref{eq:scalarPert})
or (\ref{eq:sprr}), in which the effective potential is given by
(\ref{eq:Veff}) and (\ref{eq:trans}). To calculate the QNMs, we
should impose ingoing boundary condition near the event horizon and
outgoing boundary condition near the spatial infinity. From (\ref{eq:wt})
and (\ref{eq:scalarPert}), this implies the scalar perturbation has
asymptotical behaviors as follow: 
\begin{equation}
\psi\propto\begin{cases}
e^{-i\omega r_{*}}\to(r-r_{h})^{\frac{i\omega r_{h}}{\left(\frac{Q^{2}}{f(\phi_{h})r_{h}^{2}}-1\right)\alpha(r_{h})}}, & r_{*}\to-\infty\ (r\to r_{h}),\\
e^{i\omega r_{*}}\to e^{i\omega r}r^{2iM\omega}, & r_{*}\to\infty\ (r\to\infty).
\end{cases}
\end{equation}
Here we have used the tortoise coordinate $\frac{dr_{*}}{dr}=\frac{1}{(1-\zeta^{2})\alpha}$,
and its asymptotical behaviors
\begin{equation}
r_{*}\to\begin{cases}
\frac{-r_{h}}{\left(\frac{Q^{2}}{f(\phi_{h})r_{h}^{2}}-1\right)\alpha(r_{h})}\ln(r-r_{h}), & r\to r_{h},\\
r+2M\ln r, & r\to\infty.
\end{cases}
\end{equation}

\subsubsection{Direct integration method for unstable modes}

For an unstable mode $\omega=\omega_{R}+i\omega_{I}$ with positive
imaginary part, it is easy to see that 
\begin{equation}
\psi\to\begin{cases}
A_{+}e^{-(i\omega_{R}-\omega_{I})r_{*}}\to0, & r_{*}\to-\infty\ (r\to r_{h}),\\
A_{+}e^{(i\omega_{R}-\omega_{I})r_{*}}\to0, & r_{*}\to\infty\ (r\to\infty).
\end{cases}
\end{equation}
We adopt a scheme that has been previously used to calculate the unstable
mode \cite{Blazquez-Salcedo:2018jnn,Blazquez-Salcedo:2020nhs,Liu:2022fxy}.
We introduce a squared frequency variable, $w=\omega^{2}$, and treat
it as an auxiliary function, $w(r)$, which satisfies $\frac{dw}{dr}=0$.
We solve this equation along with (\ref{eq:sprr}). The three boundary
conditions are $\psi|_{r=r_{h}}=\psi|_{r=\infty}=0$, and $\psi|_{r_{m}}=1$,
where $r_{m}$ is typically about $6r_{h}$. In practice, we set the
inner boundary at $r=r_{h}(1+10^{-6})$ and outer boundary at $r=200r_{h}$.
By following this procedure, we can successfully obtain the unstable
modes. However, this procedure is not suitable for calculating stable
modes with a negative imaginary part. 

\subsubsection{The first-order WKB method for stable modes }

To calculate the stable modes, we employ the first-order WKB method
\cite{Berti:2009kk,Konoplya:2011qq}. Using the perturbation equation
(\ref{eq:scalarPert}), the quasinormal modes are determined by the
equation 
\begin{equation}
0=\omega^{2}-V_{\text{eff}}(r_{m})+\left(\frac{1}{2}+n\right)i\sqrt{-2\frac{\partial^{2}V_{\text{eff}}(r_{m})}{\partial r_{*}^{2}}},\ \ \ n=0,1,2,...
\end{equation}
where $r_{m}$ corresponds to the location where $V_{\text{eff}}(r)$
reaches its maximum, $n$ represents the overtones, and $\frac{\partial}{\partial r_{*}}=(1-\zeta^{2})\alpha\frac{\partial}{\partial r}$.
We primarily focus on the dominant mode with $n=0$. Once we obtain
the static background solutions, we can solve above equation to determine
the dominant mode. However, it should be noted that the WKB method
is only applicable to stable modes \cite{Konoplya:2019hlu}. 

\subsubsection{Shooting method for QNMs}

In addition to the two perturbative methods we have discussed, we
will also utilize the shooting method to calculate the QNMs. Unlike
the previous methods, the shooting method is applicable for both unstable
and stable QNMs. To apply the shooting method, we first expand all
the background functions around the black hole horizon using a high-order
expansion with a parameter denoted as $n_{b}$, typically set to around
12. The expansions are given by:
\begin{equation}
h=\sum_{n=0}^{n_{b}}h_{n}(r-r_{H})^{n}.\label{eq:hH}
\end{equation}
where $h$ represents the variables $\zeta,\alpha,\phi$. By utilizing
the background equations (\ref{eq:sar}-\ref{eq:sprr0}) and providing
numerical values for $\beta,Q$, and $r_{H}$, we can efficiently
determine the coefficients $h_{n}$ order by order. Next, we expand
all the background functions near spatial infinity as follows:
\begin{equation}
\phi=\sum_{n=1}^{n_{b}}\frac{\phi_{n}}{r^{n}},\ \ \ \zeta=\sqrt{\frac{2M}{r}}\sum_{n=0}^{n_{b}}\frac{\zeta_{n}}{r^{n}},\ \ \ \alpha=\sum_{n=0}^{n_{b}}\frac{\alpha_{n}}{r^{n}},\label{eq:infn}
\end{equation}
where $\phi_{0}=Q_{s},\zeta_{0}=1$, and $\alpha_{0}=1$ to be coincident
with (\ref{eq:zpinf},\ref{eq:ainf}). We can also efficiently obtain
the coefficients $\phi_{n},\zeta_{n},\alpha_{n}$ order by order. 

Now let us consider the perturbation field $\psi$. Near the black
hole horizon and spatial infinity, we respectively assume the ingoing
and outgoing asymptotic behavior in the form:
\begin{equation}
\psi_{-}(r)=e^{-i\omega r_{*}}Y(r),\ \ \ \psi_{+}(r)=e^{i\omega r_{*}}Z(r),\label{eq:YZ}
\end{equation}
where $Y$ and $Z$ are regular functions near the horizon and infinity,
respectively. By substituting (\ref{eq:YZ}) into (\ref{eq:sprr}),
after dropping the factor $e^{\pm i\omega r_{*}}$, we obtain a linear
differential equation for $Y$ and $Z$, respectively. We expand $Y$
around the horizon and $Z$ near infinity with an order denoted as
$n_{p}$, typically chosen to be smaller than $n_{b}$:
\begin{equation}
Y=\sum_{n=0}^{n_{p}}Y_{n}(r-r_{H})^{n},\ \ \ Z=\sum_{n=0}^{n_{p}}Z_{n}r^{-n}.\label{eq:YZn}
\end{equation}
where we take $n_{p}=8$. Substituting (\ref{eq:YZn}) into the linear
differential equation for $Y$ and $Z$, and combining the results
with static background expansion coefficients (\ref{eq:hH}) and (\ref{eq:infn}),
we can efficiently determine $Y_{n}$ and $Z_{n}$ order by order,
respectively. These coefficients depend on $\omega$. Providing an
initial guess value of $\omega$, we obtain $Y(r)$ by integrating
the linear differential equation for $Y$ from the inner boundary
at $r_{b1}=r_{H}(1+10^{-3})$ to the midpoint $r_{m}\sim5r_{H}$ for
the boundary conditions $Y(r_{b1}),Y'(r_{b1})$. With the same initial
guess value of $\omega$, we obtain $Z(r)$ by integrating the linear
differential equation for $Z$ from the outer boundary at $r_{b2}\sim100r_{H}$
to the midpoint $r_{m}$ for the boundary conditions $Z(r_{b2}),Z'(r_{b2})$.
At the midpoint $r_{m}$, we require $\psi_{-}(r_{m})=\psi_{+}(r_{m})$
and $\frac{d\psi_{-}}{dr}(r_{m})=\frac{d\psi_{+}}{dr}(r_{m})$, which
leads to the condition:
\begin{equation}
\left.\frac{1}{Y}\frac{dY}{dr}-\frac{1}{Z}\frac{dZ}{dr}-2i\omega\frac{dr_{*}}{dr}\right|_{r=r_{m}}=0.
\end{equation}
If above equation is not satisfied, we need to adjust the initial
guess value $\omega$ and repeat the integration process. Due to the
limitation of numerical accuracy, we mainly calculate the dominant
modes of the static background solutions. 

\subsubsection{The QNMs for unstable and stable solutions}

\begin{figure}[h]
\begin{centering}
\begin{tabular}{cc}
\includegraphics[width=0.48\linewidth]{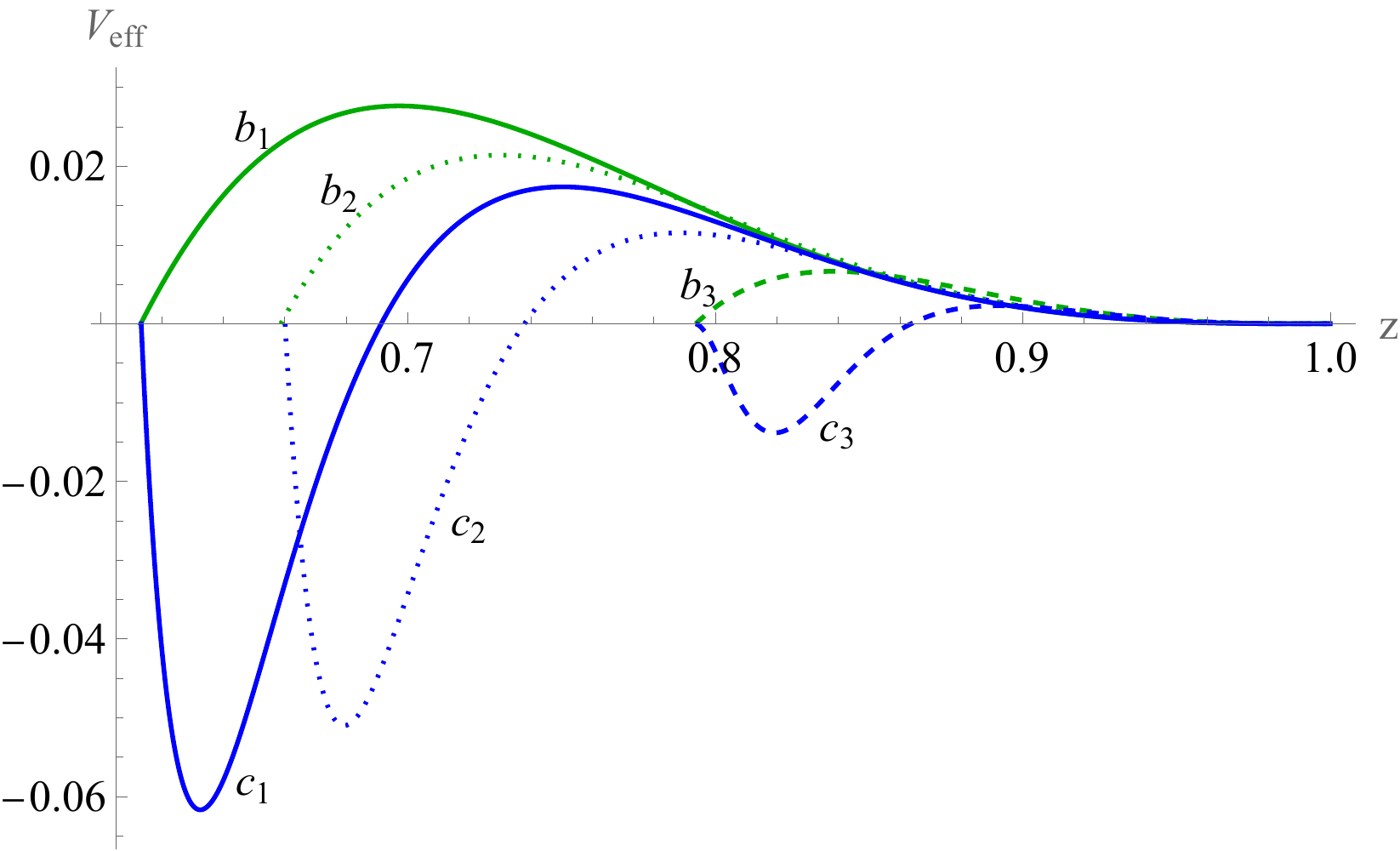} & \includegraphics[width=0.48\linewidth]{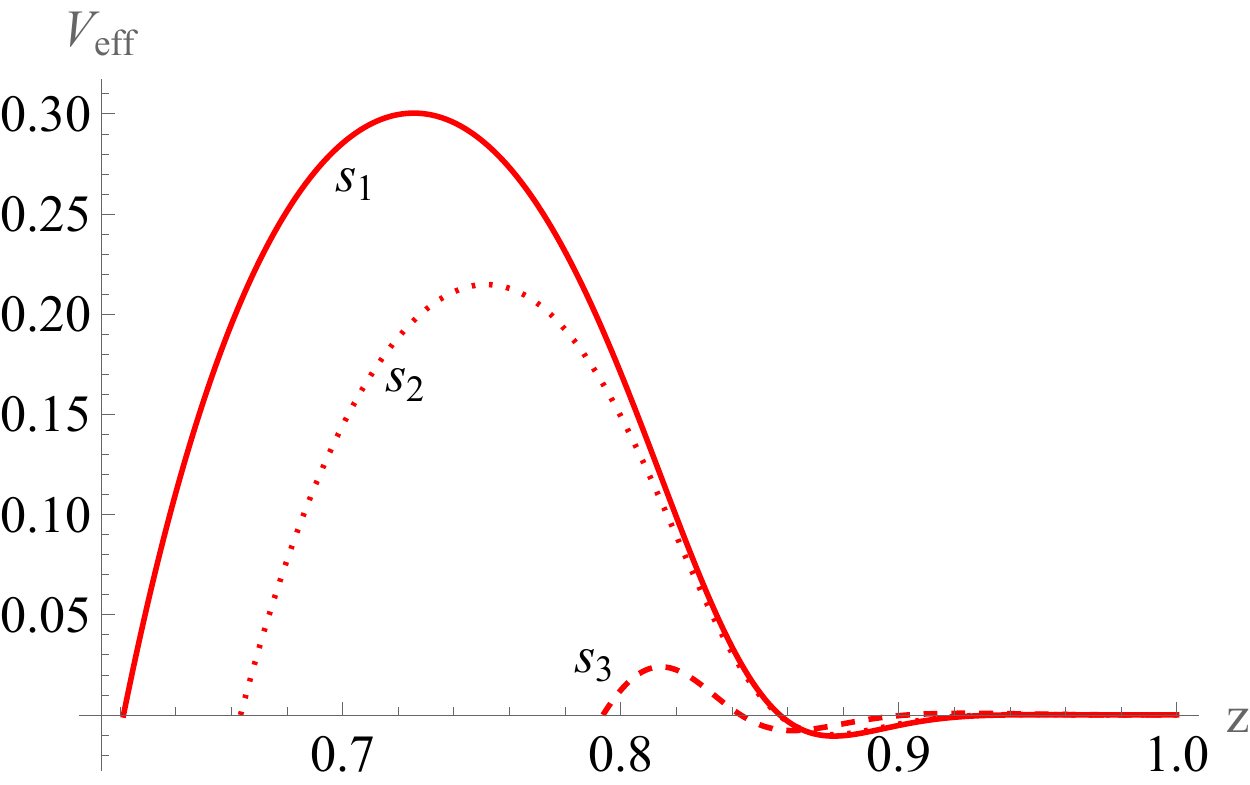}\tabularnewline
\end{tabular}
\par\end{centering}
{\footnotesize{}\caption{{\footnotesize{}\label{fig:PotentialQNM}The profiles of the effective
potential of the static solutions. The blue, green and red curves
correspond to the cold SBH $c_{1,2,3}$, RN black hole $b_{1,2,3}$
(left), and hot SBH $s_{1,2,3}$ (right), respectively. }}
}{\footnotesize\par}
\end{figure}

In Fig.\ref{fig:PotentialQNM}, we present the profiles of the effective
potential for the critical solutions $c_{1,2,3}$, bald RN black holes
$b_{1,2,3}$ and hot SBHs $s_{1,2,3}$. The BBHs and hot SBHs exhibit
potential barriers near the horizon, whereas the CSs display potential
wells in proximity to the horizon. The hot SBHs also possess potential
wells, albeit situated away from the horizon. As the total mass increases,
both the potential barrier and well diminish in size. 

According to a well known result in quantum mechanics \cite{Nandi:1995gxo},
for a perturbation equation in the form of (\ref{eq:scalarPert}),
the effective potential must satisfy the condition $\int_{-\infty}^{\infty}V_{\text{eff}}dr_{*}=\int_{r_{H}}^{\infty}\frac{V_{\text{eff}}}{(1-\zeta^{2})\alpha}dr<0$
to induce instability. We have verified that only the CSs fulfill
this condition, indicating their linear instability. Consequently,
the CSs decay either into BBHs or hot SBHs. As previously mentioned,
the CSs essentially represent cold SBHs. When they decay into BBHs,
most of their scalar hair is absorbed by the central black hole, with
only a small amount of energy escaping to infinity. On the other hand,
when the CSs decay into SBHs, a portion of the energy from the Maxwell
field is converted into scalar hair. Some of this energy dissipates
into infinity, while the remainder is absorbed by the central black
hole, causing significant growth of black hole irreducible mass and
eventual stabilization as a linearly stable hot SBH. This stabilization
is made possible by the existence of an effective potential well outside
the effective potential barrier.

\begin{table}[h]
\begin{centering}
\begin{tabular}{|cccc|ccc|}
\hline 
 & $c_{1}$ & $b_{1}$ & $s_{1}$ & $c_{2}$ & $s_{2}$ & $b_{2}$\tabularnewline
\hline 
{\footnotesize{}Direct} & {\footnotesize{}$0.132i$} &  &  & {\footnotesize{}$0.119i$} &  & \tabularnewline
\hline 
{\footnotesize{}WKB} &  & {\footnotesize{}$0.191-0.093i$} & {\footnotesize{}$0.561-0.119i$} &  & {\footnotesize{}$0.477-0.115i$} & {\footnotesize{}$0.169-0.084i$}\tabularnewline
\hline 
{\footnotesize{}Shooting} & {\footnotesize{}$0.136i$} & {\footnotesize{}$0.125-0.098i$} & {\footnotesize{}$0.548-0.129i$} & {\footnotesize{}$0.123i$} & {\footnotesize{}$0.460-0.126i$} & {\footnotesize{}$0.106-0.089i$}\tabularnewline
\hline 
{\footnotesize{}Dynamics} & {\footnotesize{}$0.136i$} & {\footnotesize{}$0.115-0.099i$} & {\footnotesize{}$0.532-0.126i$} & {\footnotesize{}$0.123i$} & {\footnotesize{}$0.432-0.116i$} & {\footnotesize{}$0.098-0.090i$}\tabularnewline
\hline 
\end{tabular}
\par\end{centering}
\caption{{\footnotesize{}\label{tab:QNMBCH}The dominant modes for the bald
RN black holes $b_{1,2}$, critical solutions or cold SBHs $c_{1,2}$,
and hot SBHs $s_{1,2}$, obtained using four different methods. The
direct integration method captures only the unstable modes, while
the WKB method provides the stable modes exclusively. The dynamical
results are extracted by using Prony method. }}
\end{table}

Table \ref{tab:QNMBCH} provides the dominant modes of the CSs $c_{1,2}$,
the final BBHs $b_{1,2}$ and the final hot SBHs $s_{1,2}$. The results
for $c_{3},b_{3}$ and $s_{3}$ exhibit similar qualitative features.
Four different methods were employed to obtain these modes, and the
results are consistent within the acceptable error range. It is noteworthy
that both the BBHs and hot SBHs consistently display complex modes
with negative imaginary parts. Conversely, the CSs consistently possess
purely positive imaginary eigenvalues. This observation bears resemblance
to the tachyonic instability exhibited by the scalar perturbation
in the context of spontaneous scalarization models \cite{Doneva:2022ewd}.  Indeed, unlike superradiant instability \cite{Zhang:2014kna,Liu:2020evp,Zhang:2020sjh}, the existence of an effective potential well near the CS horizon implies a negative effective mass squared for scalar perturbations, indicating the possibility of tachyonic instability in the CS \cite{Silva:2017uqg,Myung:2018vug,Xiong:2022ozw}.

\begin{figure}[h]
\begin{centering}
\includegraphics[width=0.5\linewidth]{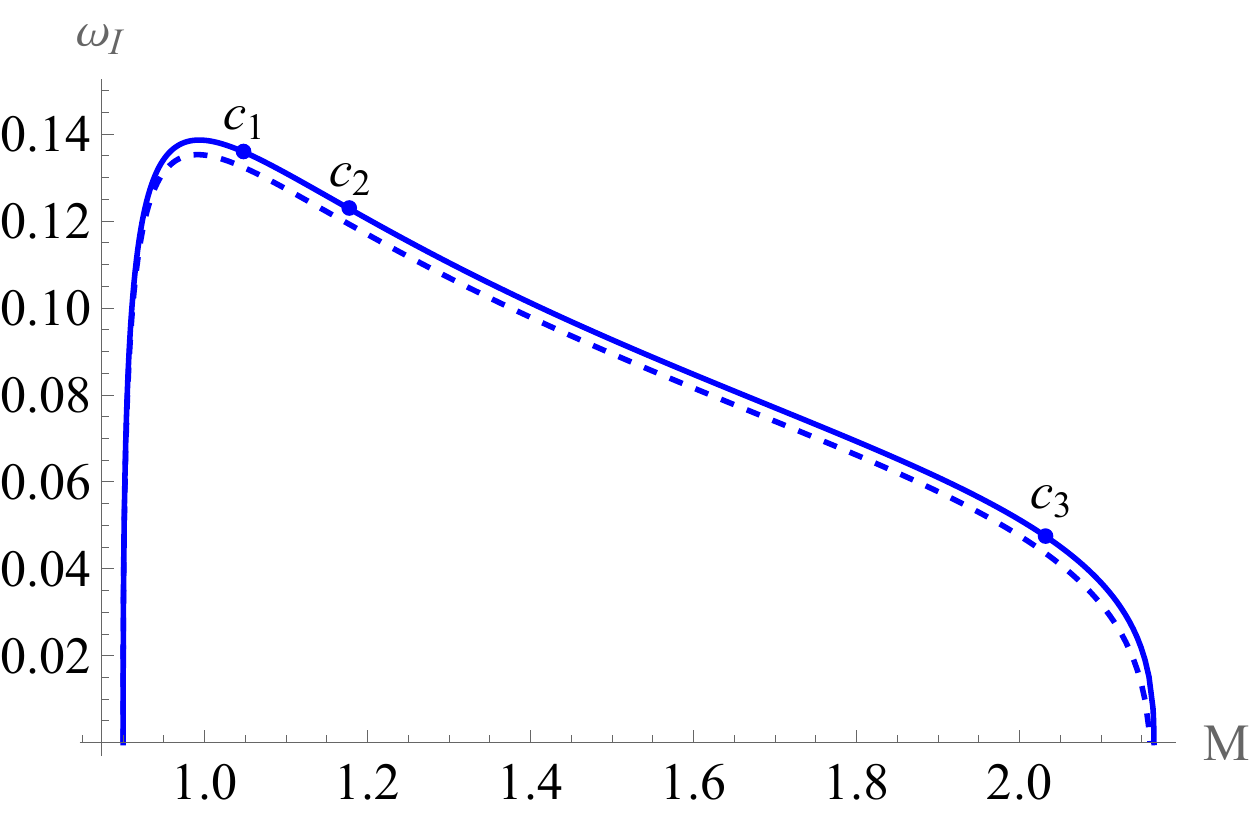}
\par\end{centering}
{\footnotesize{}\caption{{\footnotesize{}\label{fig:QNMCS}The unstable modes of the cold SBHs.
The solid curve represents the results obtained using the shooting
method, while the dashed curve corresponds to the direct integration
method. The unstable modes of the critical solutions $c_{1,2,3}$
are indicated for comparison.}}
}{\footnotesize\par}
\end{figure}

Considering the pivotal role of the CSs in the critical dynamical
transition, which are essentially the cold SBHs, we present the dominant
modes of all the cold SBHs in Fig.\ref{fig:QNMCS}. These modes are
obtained using two different methods. The cold SBH branch exhibits
an unstable mode characterized by $\omega_{R}=0$ and $\omega_{I}>0$
across its entire range of existence. At the two endpoints of the
cold SBH branch, the value of $\omega_{I}$ approaches zero. However,
it attains a maximum value at a specific $M$. We recommend referring
to \cite{LuisBlazquez-Salcedo:2020rqp} for a more comprehensive examination
of the QNMs of hot, cold and bald black holes in a similar EMS model. 

\section{Summary and Discussion \label{sec:Summary}}

The EMS models offer a fascinating theoretical framework for exploring
the intricate relationship between bald and hairy black holes. A classification
of these models based on the black hole solutions characterized by
the coupling function $f(\phi)$ was proposed in \cite{Astefanesei:2019pfq}.
Class I models exclusively accommodate hairy black hole solutions,
such as the Einstein-Maxwell-dilaton model. On the other hand, Class
IIA and IIB models allow for both RN solutions and scalarized solutions.
In Class IIA models, the scalarized branch is smoothly connected to
the RN branch, and it may exhibit tachyonic instability, resulting
in spontaneous scalarization. The model investigated in this paper
belongs to Class IIB, where three distinct solution branches coexist
within certain parameter ranges. Specifically, the RN and hot scalarized
branches are linearly stable, while the cold scalarized branch is
linearly unstable \cite{LuisBlazquez-Salcedo:2020rqp}. Notably, the
hot scalarized branch is not continuously connected to the RN branch.
The phenomenon of three branch coexistence in black hole solutions
has also been observed in scalar-Gauss-Bonnet theories \cite{Doneva:2021tvn,Blazquez-Salcedo:2022omw,Doneva:2022byd,Doneva:2022yqu,Staykov:2022uwq,Liu:2022fxy,Lai:2023gwe}.
It is important to highlight that the static or perturbative analysis
is insufficient to determine the nonlinear dynamics. In our previous
works \cite{Zhang:2021nnn,Zhang:2022cmu,Liu:2022fxy}, we first revealed
the intriguing dynamical transitions among these three solution branches.
We discovered a novel physical mechanism involving nonlinear accretion
that leads to black hole scalarization beyond spontaneous scalarization.

In this paper, we begin with the linearly stable bald RN black hole
as the initial black hole. Only when this seed black hole accretes
a sufficient amount of scalar perturbation does it undergo a transformation
into a hot SBH. Interestingly, we observe the emergence of a linearly
unstable CS precisely at the scalarization threshold. It is worth
noting that if the perturbation is excessively strong, the final solution
reverts back to being a RN black hole once again. In the context of
nonlinear dynamical evolution, a similar linearly unstable CS persists
at the threshold. From the perspective of static solutions, this behavior
is qualitatively reasonable. When the total mass of the system is
large and the charge is fixed, only RN solutions exist, while scalarized
solutions only emerge when the total mass of the system is small.
Additionally, we also consider the linearly stable hot SBH as the
initial black hole. In this case, descalarization occurs when the
scalar perturbation reaches a significant magnitude. Similarly, we
observe the appearance of a linearly unstable CS at the descalarization
threshold. We have analyzed effective potentials of the perturbation
for all the CSs, and conclude that the CSs suffer tachyonic instability,
just as the RN black holes suffer tachyonic instability under small
scalar perturbation in the class IIA models. So the nonlinear scalarization
of the bald black holes are induced by the tachyonic instability of
the CSs. These results are not specific to our chosen initial configurations,
but rather exhibit a qualitatively universal nature. We have further
investigated other coupling functions belonging to Class IIB, such
as $f(\phi)=e^{\beta\phi^{n}}$ or $1+\beta\phi^{n}$ when $n\geq3$,
and consistently observed qualitatively similar critical dynamical
behaviors at the scalarization and descalarization thresholds. 

The critical behaviors observed in dynamical scalarization and descalarization
at the thresholds bear resemblance to the type I critical gravitational
collapse. When the perturbation parameter $p$ approaches the threshold
$p_{*}$, all intermediate solutions are attracted to a linearly unstable
CS and remain on this CS for a duration described by $T\propto-\gamma\ln|p-p_{*}|$.
Here the coefficient $\gamma$ equals the reciprocal of the eigenvalue
of the single unstable mode associated with the CS. In type I critical
gravitational collapse, the CS is unique, resulting in a universal
$\gamma$ value that applies to all families of initial data. However,
in the case of type I dynamical scalarization and descalarization,
the CSs are not unique. Specifically, the CS belongs to the cold SBH
branch, which exists within a certain parameter range. Consequently,
for different initial data families, the intermediate solutions will
be attacted to different CSs, leading to varying values of the corresponding
coefficient $\gamma$. Furthermore, we discover that the irreducible
masses of the final black holes in critical dynamical scalarization
and descalarization follow power laws with fractional indices, a characteristic
absent in type I critical gravitational collapse. These indices also
depend on the initial data family. We propose that these fractional
power laws arise due to the energy of matter escaping to infinity.
However, a more comprehensive investigation is necessary to delve
into this matter in the future. 

The type I critical dynamcial scalarization and descalarization have
been studied in the EMS theories \cite{Zhang:2021nnn,Zhang:2022cmu}
as well as in a generalized scalar-tensor theory \cite{Liu:2022fxy}.
Interestingly, it appears that dynamical descalarization consistently
occurs before the junction point where the cold SBH and hot SBH branches
meet in EMS theories. However, in the context of the generalized scalar-tensor
theory, dynamical descalarization consistently takes place precisely
at the junction point. The underlying cause for this discrepancy remains
unknown at present. We speculate that this distinction could be attributed
to the coupling between either the gravitational curvature invariant
or the matter invariant. To confirm this conjecture, further investigations
regarding type I critical dynamics of black holes in other models
are necessary.

Finally, let us explore some potential extensions of this work. The
type I critical dynamical behaviors we have uncovered are not limited
to scalarization and descalarization phenomena alone. In our preliminary
research, we have found indications that these critical dynamics also
manifest in black hole transitions involving Q-hair \cite{Herdeiro:2020xmb,Hong:2020miv},
and a dedicated paper on this topic is forthcoming. Moreover, when
examining the phase diagram presented in Fig.\ref{fig:static}, we
observe certain resemblances to those observed in higher-dimensional
spacetimes for black rings and Myers-Perry black holes \cite{Emparan:2008eg}.
This prompts us to speculate the existence of a type I critical dynamical
transition bridging these systems. Furthermore, there are indications
that type I critical dynamics manifest in the real-time processes
of phase separation and nucleation during holographic first-order
phase transitions \cite{Attems:2019yqn,Bea:2020ees,Bea:2022mfb,Janik:2017ykj,Janik:2022wsx},
which have been examined in studies such as \cite{Chen:2022cwi,Chen:2022tfy}.
Additionally, critical dynamics should exist in the first-order phase
transitions of matter within specific gravity systems, resulting in
the emission of gravitational waves \cite{Bauswein:2018bma,Most:2018eaw,Weih:2019xvw}.
Overall, the investigation of type I critical dynamics holds immense
theoretical and experimental significance across various domains.
These potential extensions shed light on the far-reaching implications
and broad applicability of type I critical dynamics in diverse physical
phenomena.

\section*{Acknowledgments}

This work is supported by Natural Science Foundation of China (NSFC)
under Grant No. 11975235, 12005077, 12035016 and 12075202, and Guangdong
Basic and Applied Basic Research Foundation under Grant No. 2021A1515012374. 

\bibliographystyle{apsrev4-2ideal}
\bibliography{EMSCriticalDetailed}

\end{document}